\UseRawInputEncoding
\documentclass[twocolumn]{aastex7}
\usepackage{CJK}
\usepackage{amsmath}
\usepackage{graphicx}
\usepackage{breqn}

\newcommand{\ntois}{79 }

\newcommand{\teffsp}{\ensuremath{T_{\textrm{eff} } } }

\newcommand{\avgenosp}{\ensuremath{\langle e \rangle}}
\newcommand{\avge}{\ensuremath{\langle e \rangle } }

\newcommand{\nplanets}{\ensuremath{236} }

\defcitealias{sagear_orbital_2023}{S23}
\defcitealias{gilbert_planets_2025}{G25}

\begin{document}

\title{The Orbital Eccentricity--Radius Relation for Planets Orbiting M Dwarfs}

\author[0000-0002-3022-6858]{Sheila Sagear}
\affiliation{Department of Astronomy, University of Florida, Gainesville, FL 32611}
\email[show]{ssagearastro@gmail.com}

\author[0000-0002-3247-5081]{Sarah Ballard}
\affiliation{Department of Astronomy, University of Florida, Gainesville, FL 32611}
\email{sarahballard@ufl.edu}

\author[0000-0003-0742-1660]{Gregory J. Gilbert}
\affiliation{Department of Astronomy, California Institute of Technology, Pasadena, CA, USA}
\affiliation{Department of Physics and Astronomy, University of California, Los Angeles, USA}
\email{ggilbert@caltech.edu}

\author[0009-0008-1188-2025]{Mariangel Albornoz}
\affiliation{Department of Astronomy, University of Florida, Gainesville, FL 32611}
\email{malbornoz@ufl.edu}

\author[0000-0002-1910-5641]{Christopher Lam}
\affiliation{Department of Astronomy, University of Florida, Gainesville, FL 32611}
\email{c.lam@ufl.edu}

\begin{abstract}

The orbital eccentricity-radius relation for small planets is indicative of the predominant dynamical sculpting processes during late-stage orbital evolution. Previous studies have shown that planets orbiting Sun-like stars exhibit an eccentricity-radius trend such that larger planets have higher orbital eccentricities, and that radius gap planets may have modestly higher orbital eccentricities than planets on either side of the radius gap. In this work, we investigate the trend for a sample of smaller M dwarf stars. For a sample of \nplanets single-- and multi--transit confirmed planets or candidates discovered by the \textit{TESS} and \textit{Kepler} missions, we constrain orbital eccentricity for each planet from the transit photometry together with a stellar density prior. We investigate the binned eccentricity-planet radius relation for the combined planet sample and present evidence for a positive eccentricity-radius relationship with elevated eccentricities for planets larger than $3.5$ $R_{\oplus}$, similar to the trend for planets orbiting Sun-like stars. We find modest evidence that single-transit M dwarf planets near the radius gap exhibit higher eccentricity, consistent with trends for Sun-like stars. However, we see no evidence for an increased eccentricity near the radius gap among multi-transit M dwarf planets. We discuss implications for these results in the context of predominant atmospheric loss mechanisms: namely, supporting evidence for photoevaporation in M dwarf planets vs. planet-planet collisions or giant impacts in FGK dwarf planets.

\end{abstract}


\section{Introduction} \label{sec:intro}

NASA's \textit{Kepler} mission \citep{borucki_kepler_2010} revealed that exoplanets are abundant in the Milky Way, and that the vast majority of known exoplanets are smaller than Neptune ($< 4 R_{\oplus}$). The occurrence of small planets has been shown to increase for later host stellar spectral types \citep{dressing_occurrence_2015, mulders_slar-mass-dependent_2015}. These small planets themselves exhibit a bimodal radius distribution, with a dearth of planets between $1.5$ and $2$ $R_{\oplus}$: the so-called ``radius gap" or ``radius valley" \citep{fulton_california-kepler_2017, fulton_california-kepler_2018}. The radius gap may be a function of orbital period, stellar age, stellar phase space density, or some combination of these \citep{vaneylen_asteroseismic_2018, berger_revised_2018,
berger_gaia-kepler_2020,
david_evolution_2021}. Notably, the radius gap is also a function of host star mass, with the radius gap appearing at smaller radii for lower-mass stars \citep{hardegree-ullman_kepler_2019, berger_gaia-kepler_2020, gaidos_radius_2024, ho_shallower_2024}.

M dwarfs are particularly interesting as planet hosts for a multitude of reasons. M dwarfs host small, terrestrial planets at higher rates than more massive stars \citep{france_high-energy_2020, hsu_occurrence_2020}. Transits of small planets around M dwarfs are also deeper than planets of the same size around FGK stars, and the large numbers of nearby M dwarfs are also uniquely suited for radial velocity follow-up observations.  For these reasons, M dwarfs are the predominant stellar host in the search for terrestrial planets in the habitable zone, given the abundance of nearby M dwarfs and conservative estimates of habitable-zone planet occurrence being in the $15-30 \%$ range \citep{dressing_occurrence_2013, hsu_occurrence_2020, kopparapu_revised_2013}. 

Investigating the eccentricity\textemdash radius relation for M dwarf planets is pressing over and above their ubiquity: these properties also play outsized roles in M dwarf planet habitability. Because the habitable zone is so close to the host star, even a modest eccentricity can produce a runaway ``tidal Venus" effect \citep{barnes_tidal_2013}. The proximity to the star also means that habitable-zone planets lie inside the nominal spin-synchronization radius, so that their size plays a role in atmospheric stability \citep{heng_stability_2012, yang_strong_2014}. The growing sample size of known M dwarf planets enables studies of their dynamics, whether through transits themselves or through radial velocity follow-up. The \textit{TESS} mission has been particularly pivotal for characterizing small planets around nearby cool stars, and the mission has recently expanded the catalogue of known M dwarf planets to several hundreds \citep{guerrero_tess_2021}. 

We have in hand now a sizable sample of M dwarf planets, sufficient to consider how orbital dynamics various across spectral type. For FGK dwarfs, \citet{van_eylen_orbital_2019} found that \textit{Kepler} planets with $R_p < 6 R_{\oplus}$ have generally low eccentricities (\avge $\sim 0.1-0.3$). \citet{sagear_orbital_2023} (hereafter \citetalias{sagear_orbital_2023}) found that the underlying orbital eccentricity distribution for small planets around M dwarfs is remarkably similar, with \avge $\sim 0.03-0.2$. \citet{gilbert_planets_2025} (hereafter \citetalias{gilbert_planets_2025}) refined this trend toward low eccentricities for small planets, finding \avge $\sim 0.05$ for planets smaller than $\sim 3.5$ $R_{\oplus}$ and a sharp transition to higher eccentricities \avge $\sim 0.2$ for larger planets. These results suggest that small and large planets form via distinct channels, but that formation histories may be common across a range of stellar types. However, eccentricity information is just one axis of several that are required to determine the likeliest formation and evolution mechanisms for planets around different spectral types of host stars. Considering orbital dynamical information in tandem with radii may reveal deeper trends in the architectures of M dwarf and FGK dwarf systems, and point to certain formation and evolution mechanisms for each. Understanding the eccentricity--radius relation for M dwarfs contributes to a more complete picture of formation and atmospheric loss mechanisms for typical planets in the Milky Way. 

The radius distribution may be intrinsically imprinted during formation. Alternatively, some mechanism(s) leading to atmospheric loss may be the driving force for the radius gap. Key atmospheric loss mechanisms include photoevaporation, whereby X-ray and extreme ultraviolet radiation is sufficient for the atmosphere to escape \citep{owen_planetary_2012}, and core-powered mass loss, where remnant thermal energy from a planet's cooling core plays a large role \citep{ginzburg_super-earth_2016, rogers_photoevaporation_2021}. In planetary evolution models, both stellar-luminosity-driven mass loss mechanisms (photoevaporation and core-powered mass loss) reproduce the radius gap for small planets \citep{lopez_role_2013, owen_kepler_2013, jin_planetary_2014, chen_evolutionary_2016, ginzburg_core-powered_2018, gupta_sculpting_2019, gupta_signatures_2020}. Indeed, both atmospheric loss mechanisms can dominate in different regimes depending on the location of the extreme ultraviolet penetration depth (a parameter related to the planet's radius), and may even occur concurrently \citep{owen_mapping_2024}. Alternatively, atmospheric loss may also be catalyzed by late-stage giant impacts \citep{liu_suppression_2015, schlichting_atmospheric_2015, schlichting_atmosphere_2018, biersteker_atmospheric_2019, chance_signatures_2022}, in which a planetary embryo impacts a planet with a primordial atmosphere leading to heating of the planet's interior and hydrodynamic ejection of the atmosphere, leading to atmospheric loss. 

We expect the dominant mechanism for atmospheric loss in small planets to leave some demographic signature on planetary system architectures and orbital dynamics. For example, modeling photoevaporation and giant impacts through planet formation models results in different numbers of planets in the radius gap (photoevaporation leads to a nearly empty radius gap, while giant impacts leads to a non-empty radius gap) \citep[e.g.][]{venturini_nature_2020,izidoro_exoplanet_2022, owen_mapping_2024, ho_shallower_2024}. Furthermore, \citet{chance_evidence_2024} show through atmospheric loss simulations that if giant impacts are the predominant driver of atmospheric loss, planets in the radius gap should exhibit higher eccentricities than planets with lower and higher radii. Indeed, \citetalias{gilbert_planets_2025} observed that small planets around FGK dwarfs in the radius gap exhibit modestly higher eccentricities than their neighbors on either side of the radius gap.

In this work, we examine the extent to which the   \citetalias{gilbert_planets_2025} finding, focused upon on the eccentricity\textemdash radius relation of planets orbiting FGK dwarfs, also applies to planets orbiting M dwarfs. We investigate two main scientific questions:

\begin{enumerate}

\item For M dwarf planets, is there a transition to higher eccentricities beginning at $\sim 3.5$ $R_{\oplus}$?

\item Are orbital eccentricities elevated for M dwarf planets near the radius valley?

\end{enumerate}

For both planetary radii and eccentricity measurements in this work, we rely upon transit photometry. While the gold standard for individual orbital eccentricity measurements is radial velocity (RV) data, transits also encode information (albeit coarser information) about eccentricity. \cite{ford_characterizing_2008}, \cite{kipping_novel_2012}, and \cite{dawson_photoeccentric_2012} originally demonstrated the extraction of eccentricity posteriors from transits with a technique called the ``photoeccentric effect". The photoeccentric effect formalism is fully described in \citet{dawson_photoeccentric_2012}, but we briefly summarize it here. This technique leverages known information about the orbital mechanics of a planetary system along with transit observables, such as the transit duration and ingress/egress duration, to make inferences about the speed of the planet in transit and thus, the orbital eccentricity. 

This manuscript is organized as follows. In Section \ref{sec:data}, we introduce the sample of planets employed in this work. In Section \ref{sec:analysis}, we discuss the statistical methods we use to investigate the eccentricity\textemdash radius relationship. In Section \ref{sec:results}, we introduce our results and their significance. We discuss the implications of our results, including the extent to which we can answer questions (1) and (2), in Section \ref{sec:discussion}. We conclude in Section \ref{sec:conclusions}.

\section{Data} \label{sec:data}

\subsection{Planet Sample} \label{subsec:planetsample}

Our sample included transiting exoplanets discovered by both NASA's \textit{Kepler} and \textit{TESS} missions. Our \textit{Kepler} sample, originally published in \citetalias{sagear_orbital_2023}, comprised 163 planets. We identified M dwarf hosts with the criteria described in \citetalias{sagear_orbital_2023}, which included taking all KOI hosts with $T_{\mathrm{eff}}< 4000K$. Since that time, the sample of known M dwarf exoplanets has grown to include an increasing number discovered with NASA's \textit{TESS} Mission. It is not immediately obvious that the precision of eccentricity measurement from a typical \textit{TESS} lightcurve is comparable to that of \textit{Kepler}-- both the duration of transit observations and the photometric precision differ between these missions. However, \textit{TESS} data has proven to enable sufficiently informative eccentricity measurements for demographic analysis \citep[e.g.,][]{kipping_near-circular_2025, fairnington_eccentricity_2025}. A brief analysis comparing the precision of \textit{Kepler} and \textit{TESS} eccentricity measurements is presented in Appendix \ref{appendix:validation}.

We add \ntois \textit{TESS} Objects of Interest (TOIs) to our sample, identifying them as follows. We include all TOIs around M dwarf host stars with \teffsp $< 4000$ K and log($g$) $> 4.4$ according to the \textit{TESS} Input Catalog (TIC) \citep{stassun_tess_2018}. We include only TOIs with dispositions of ``Confirmed Planet" (CP) or ``Planet Candidate" (PC). Motivated by the idea that planet candidates around multi-planet systems are significantly less likely to be caused by astrophysical false positives compared to single-transit candidates \citep{ragozzine_value_2010, lissauer_closely_2011, latham_first_2011}, we include all multi-transit TOI systems where at least one member has a disposition of ``PC". We only include the single-planet systems with a disposition of ``CP". While we cannot rule out false positives for planet candidates, these selection criteria allow us to maintain a robust sample of transiting planets while still maintaining a sufficiently large sample size.

For ease of comparison with \cite{gilbert_planets_2025}, we adopt their orbital period criteria for inclusion. Thus, from both the \textit{Kepler} and \textit{TESS} samples, we remove planets with orbital periods greater than 100 days and less than 1 day. We remove KOI 950.01 due to a potentially unreliable transit fit. We were unable to include several TOIs due to computational resource limitations or incomplete data availability; these include TOIs 540.01, 700.01, 704.01, 1746.01, 1746.02, 4201.01, and 5205.01, and 5293.01. Various TOIs were excluded due to failure to find a sufficiently convergent transit fit solution. These include single-transit TOIs 237.01, 620.01, 704.01, 833.01, 910.01, 1231.01, 1470.01, 1640.01, 1696.01, 2221.01, 3984.01, 4184.01, and multi-transit TOIs 732.02, 1468.02, 2084.01, 2267.02, 6022.01, along with systems 1224, 2269, and 5523. The complete sample of TOIs used in this analysis is listed in Table \ref{tab:fitplanetparams}.

The photoeccentric technique requires a density prior for the stellar host. In \citetalias{sagear_orbital_2023}, we calculated stellar density using the \citet{mann_how_2015} radius-$M_{k}$ and \citet{mann_how_2019} mass-$M_{k}$ relations, along with their \textit{Gaia} Data Release 2 (DR2) \citep{prusti_gaia_2016, brown_gaia_2018} parallax measurements. For our newly added TOIs, \citet{stassun_revised_2019} flagged cool dwarfs in the revised TIC and calculated their stellar densities using the same \citet{mann_how_2015} and \citet{mann_how_2015} empirical relationships and \textit{Gaia} parallax measurements. We therefore draw directly from \citet{stassun_revised_2019} for the stellar densities of TOI host stars. 

 \subsection{Lightcurve Preparation} \label{subsec:lightcurve_preparation}

For the treatment of the photometry from \textit{Kepler}, we refer the reader to \citetalias{sagear_orbital_2023}, and focus here upon the \textit{TESS} lightcurves. We download the \textit{TESS} lightcurves from the Mikulski Archive for Space Telescopes (MAST) using the \texttt{lightkurve} Python package \citep{lightkurve_collaboration_lightkurve_2018}. For each TOI in our sample, we download all available quarters of \textit{TESS} lightcurves, normalize them to a mean flux value of 0, and stitch the quarters together together. We sigma-clip the data to $9\sigma$, except for TOI 2119.01 where we sigma-clip to $15\sigma$ to ensure none of the in-transit data was removed.

We calculate the expected transit times for each TOI using the published period ($P$) and transit epoch ($t_0$) values in the \textit{TESS} Input Catalog. We remove the out-of-transit data (data points more than 6 hours before or after each expected transit midpoint). For known multi-transit systems, we also remove all simultaneous transits with midpoints that lie within 6 hours of another transit midpoint in the same system. This leaves us with individual transit segments, padded by approximately 6 hours of out-of-transit data before and after the transit, for each TOI.

To remove the effect of long-term trends, we mask the in-transit data using the expected transit midpoint and transit duration published the TIC. For each individual transit segment, we fit a line to the data immediately before and after the transit using the \texttt{numpy.polyfit} least-squares fitting function. We interpolate the line across the entire transit segment and subtract the line from the data, flattening the transit segment. We elect to perform this line-fitting process rather than e.g. fitting and removing the noise parameters of the entire light curve primarily for computational efficiency and consistency with \citetalias{sagear_orbital_2023}, with which we strive to create a consistent sample.

\section{Analysis} \label{sec:analysis}

\subsection{Lightcurve fitting} \label{lcfitting}

We perform the lightcurve modeling with gradient descent using the \texttt{jaxoplanet} and \texttt{numpyro} Python packages \citep{hattori_exoplanet-devjaxoplanet_2024, phan_composable_2019}. The transit fit free parameters and their priors are listed in Table \ref{table:transitfitfreeparameters}. The free parameters include the  orbital period $P$, transit midpoint $t_0$, logarithm of planet radius $log(r_p)$, impact parameter $b$, re-parameterized limb darkening parameters $q_1$ and $q_2$ \citep{kipping_efficient_2013}, and logarithm of the full transit duration $log(T_{14})$. We elect to fit each planet in multi-transit systems individually for computational efficiency; since we directly sample the transit duration and only incorporate the stellar density in the post-fitting Importance Sampling steps (described in Section \ref{subsubsec:importance}), there is little risk of creating inconsistent stellar models for multiple planets in the same system. The only such risk comes from generating inconsistent limb-darkening parameter posteriors. We inspect the limb-darkening posteriors for all planets in multi systems to ensure that they are consistent; we find that the best-fit quadratic limb darkening parameters for planets within a system are all within one sigma of each other.

We initialize the sampler to the values published in the TIC. We sampled the model parameters using No U-Turn Sampling (NUTS) \citep{homan_no-u-turn_2014} with two chains of 5000 tuning steps followed by at least 5000 posterior draws each. To check the convergence of each fit, we check that the Gelman-Rubin statistic $\hat{R}$ \citep{gelman_inference_1992} for all free parameters is $<1.05$, along with the sample acceptance rate and number of effective samples for each fit.

\subsubsection{Planets with Transit Timing Variations}

Planetary systems commonly exhibit transit timing variations (TTVs) such that a planet's transit time may be shifted for each subsequent transit. Due to the short baseline observation time for TOIs outside of the Continuous Viewing Zone, there is a lack of robust TTV measurement curves for most TOIs in our sample. We fit individual transit times for TOI systems 122, 177, 269, 700, 789, 1201, 1695, 1801, 2136, 2257, 2267, 2406, 4336, 4438, and 4479 to account for potential TTVs. Though not all of these planets are known in the literature to exhibit TTVs, we liberally flag planets to have TTVs modeled. We flag planets for TTV fitting based on visual inspection of the phase-folded light curve or multimodal transit fit results using a linear ephemeris, in addition to conducting a literature search for planets with TTVs.

We fit transit lightcurves flagged for TTVs by allowing the transit time to float freely within a uniform prior of 0.1 days before and after the expected transit time. Since we allow individual transit times to float uniformly, rather than explicitly modeling the TTV curve over time, fitting transit times for planets with no TTVs should have no effect on the resulting transit fit; yet, all TTVs may not be known for all \textit{TESS} planets in our sample. Since the \texttt{jaxoplanet} package did not provide functionality to allow for free-floating transit times (at the time of analysis), we instead used the \texttt{exoplanet} and \texttt{pymc} Python packages to fit these lightcurves. The \texttt{exoplanet} package was the primary tool used to fit \textit{Kepler} transit lightcurves, including those with TTVs, in \citetalias{sagear_orbital_2023}.

We process the lightcurves for TOIs with TTVs according to the procedure described in Section \ref{subsec:lightcurve_preparation}, with the adjustment made to accommodate the variable transit times: instead of fitting a linear model to the data 6 hours before and after each expected transit time, with the duration of the transit masked, we instead fit a linear model to the data 9 hours before and after each expected transit time, with the duration+4.8 hours (0.2 days) masked. This process ensures that where TTVs are present, in-transit data will not be inadvertently included in the immediate out-of-transit data used to normalize and detrend the lightcurve.

We apply the same free parameters and priors listed in Table \ref{table:transitfitfreeparameters}, except for the $t_0$ prior. We include each individual transit time as a free parameter and sample the individual transit times uniformly between $t_0-0.1$ days and $t_0+0.1$ days, where $t_0$ is the expected transit midpoint. Using the NUTS sampler implementation in the \texttt{pymc} package, we run two chains of at least 5000 tuning steps followed by at least 5000 posterior draws each. As above, we check the convergence of each fit with the Gelman-Ruban statistic $\hat{R}$, the sample acceptance rate, and number of effective samples.

\subsubsection{Implementation of Importance Sampling} \label{subsubsec:importance}

Fitting transit lightcurves using Markov Chain Monte Carlo (MCMC) sampling methods can be prohibitively resource-intensive due to the high number of steps needed for convergence and non-standard posterior shapes. We use the method described by \citet{macdougall_accurate_2023} to avoid directly sampling most transit parameters. Instead, we generate eccentricity constraints by directly sampling the transit duration. This method results in a significant reduction in the computational resources required. We summarize this method here, which is fully described in \citet{macdougall_accurate_2023}. 

Rather than explicitly constraining $e$, $\omega$, and $\rho_{\star}$ during the sampling process, we instead parametrize the transit model in terms of the transit mid-time $t_0$, transit duration $T_{14}$, planet-to-star radius ratio $r_p/r_s$, impact parameter $b$, and triangular quadratic limb darkening parameters $q_1$ and $q_2$ (such that $q_1 = (u_1 + u_2)^{\mathbf{2}}$ and $q_2 = 0.5u_1(u_1 + u_2)^{\mathbf{-1}}$) \citep{kipping_efficient_2013}. Then, we estimate the values of $e$, $\omega$, and $\rho_{\star}$ through post-model importance sampling. 

\citet{macdougall_accurate_2023} thoroughly demonstrate that the importance sampling transit fitting procedure results in comparable results to directly sampling transit parameters, with improvements in performance and resource usage. We have performed a secondary, independent test using \textit{Kepler} M Dwarf planets to ensure that results from the two methods are comparable. We processed a \textit{Kepler} light curve for an M dwarf planet in the manner described in \citetalias{sagear_orbital_2023}, and performed the transit light curve fit using the importance sampling procedure. We compared the transit fit posteriors, including the eccentricity posteriors, to the results in \citetalias{sagear_orbital_2023} and found nearly identical results. We describe this validation process in detail in Appendix \ref{appendix:validation}.

Using the posterior samples for $t_0$, $r_p$, $b$, $u_1$ and $u_2$, and the transit duration, we use the Importance Sampling procedure described by \citet{macdougall_accurate_2023} to obtain eccentricity $e$ and longitude of periastron $\omega$ posterior samples. We first define test $e$ and $\omega$ arrays by randomly generating values uniformly between 0 and 0.95 (for $e$) and between $\frac{- \pi}{2}$ and $\frac{3 \pi}{2}$ (for $\omega$). Then, we calculate $\rho_{\rm \star, samp}$, the stellar density calculated from the transit fit posterior samples assuming a circular orbit, with
\begin{equation} \label{eq:rho_circ_to_star}
    \rho_{\rm \star, samp} = \frac{3 \pi}{G P^2}\left(\frac{\left(1+r_p\right)^2-b^2}{\sin^2{\left(\frac{T_{\rm 14} \pi}{P}\frac{1+e\sin{\omega}}{\sqrt{1-e^2}}\right)}}+b^2\right)^{3/2}.
\end{equation}
We substitute $P$ with the orbital period published in the TIC and $r_p$, $b$, and $T_{14}$ directly with the transit fit posterior samples. Therefore, we take $\rho_{\rm \star, true}$ to be the stellar density published in the TIC for each sample star (described in Section \ref{sec:data}). We then compare $\rho_{\rm \star, samp}$ with $\rho_{\rm \star, true}$ by calculating the log-likelihood of each posterior sample with

\begin{equation}
\label{eq:loglike}
    \log \mathcal{L}_i = -\frac{1}{2}\Big(\frac{\rho_{\rm \star, samp, i} - \rho_{\rm \star, true}}{\sigma_{\rho_{\rm \star, true}}}\Big)^2,
\end{equation}

We calculate and assign an importance weight to each randomly generated $e$ and $\omega$ sample using

\begin{equation}
\label{eq:ewweights}
    w_i = \frac{\mathcal{L}_i}{\sum_i \mathcal{L}_i}
\end{equation}
The importance-weighted $e$ and $\omega$ distributions comprise the final ($e, \omega$) posterior distributions for each transit fit.

\begin{deluxetable*}{cccc}
\tablecaption{Transit fit free parameters, prior distributions and values. \label{table:transitfitfreeparameters}}
\tablehead{\colhead{Free Parameter} & \colhead{Description} & \colhead{Prior Distribution} & \colhead{Prior Values}}
\startdata
$P$ & Period & Truncated Normal & $\sim \mathcal{TN} (P_{\mathrm{TOI}}, 0.0001, P_{\mathrm{TOI}}-0.005, P_{\mathrm{TOI}}+0.005)$ (days) \\ 
$t_0$ & Transit mid-time & Uniform & $\sim \mathcal{U}(t_{0, \mathrm{TOI}}-0.01, t_{0, \mathrm{TOI}}+0.01)$ (days) \\
$log(\mathrm{T_{14}})$ & Log of full transit duration & Normal & $\sim \mathcal{N} (T_{14, \mathrm{TOI}}, 0.2)$ (days) \\
$log(R_p)$ & Log of planet radius & Normal & $\sim \mathcal{N} (R_{p, \mathrm{TOI}}, 0.2)$ $(R_{\star})$ \\
$b$ & Impact parameter & Uniform & $\sim \mathcal{U}(0, 1.2)$ \\
$(q_1, q_2)$ & Triangular Limb Darkening Parameters & Uniform & $\sim \mathcal{U} [0, 1],$ $\mathcal{U} [0, 1]$ \\
\enddata
\tablecomments{The prior values with subscript ``TOI" denote the individual planet parameters published in the NASA Exoplanet Archive \textit{TESS} Project Candidates table.}
\end{deluxetable*}

\begin{deluxetable*}{ccccccc}
\tablecaption{Transit fit parameters for TOIs. \label{tab:fitplanetparams}}
\tablehead{\colhead{TOI} & \colhead{$R_p/R_{\star}$} & \colhead{$b$} & \colhead{$T_{14}$ (days)} & \colhead{$e$} & \colhead{$\omega$} & $e_{mode}$}
\startdata
122.01 & $0.076_{-0.003}^{+0.003}$ & $0.436_{-0.284}^{+0.234}$ & $0.049_{-0.002}^{+0.002}$ & $0.351_{-0.229}^{+0.251}$ & $1.374_{-1.429}^{+1.831}$ & $0.346$ \\
127.01 & $0.217_{-0.004}^{+0.003}$ & $0.378_{-0.087}^{+0.06}$ & $0.084_{-0.001}^{+0.001}$ & $0.138_{-0.091}^{+0.304}$ & $3.058_{-3.749}^{+0.728}$ & $0.055$ \\
134.01 & $0.024_{-0.001}^{+0.001}$ & $0.396_{-0.263}^{+0.273}$ & $0.049_{-0.001}^{+0.001}$ & $0.276_{-0.181}^{+0.3}$ & $1.448_{-1.619}^{+1.972}$ & $0.237$ \\
177.01 & $0.037_{-0.001}^{+0.001}$ & $0.24_{-0.171}^{+0.219}$ & $0.051_{-0.001}^{+0.001}$ & $0.404_{-0.114}^{+0.248}$ & $1.53_{-1.185}^{+1.328}$ & $0.334$ \\
198.01 & $0.027_{-0.002}^{+0.005}$ & $0.525_{-0.328}^{+0.264}$ & $0.108_{-0.006}^{+0.009}$ & $0.264_{-0.162}^{+0.304}$ & $1.934_{-2.897}^{+2.079}$ & $0.129$ \\
233.01 & $0.05_{-0.002}^{+0.002}$ & $0.608_{-0.346}^{+0.171}$ & $0.087_{-0.004}^{+0.004}$ & $0.214_{-0.144}^{+0.349}$ & $1.479_{-1.972}^{+2.333}$ & $0.129$ \\
233.02 & $0.033_{-0.004}^{+0.006}$ & $0.807_{-0.361}^{+0.121}$ & $0.063_{-0.006}^{+0.008}$ & $0.285_{-0.189}^{+0.363}$ & $1.303_{-2.094}^{+2.665}$ & $0.204$ \\
244.01 & $0.03_{-0.001}^{+0.001}$ & $0.387_{-0.259}^{+0.282}$ & $0.052_{-0.001}^{+0.002}$ & $0.454_{-0.22}^{+0.22}$ & $1.759_{-1.488}^{+1.192}$ & $0.46$ \\
269.01 & $0.065_{-0.003}^{+0.002}$ & $0.557_{-0.298}^{+0.143}$ & $0.038_{-0.001}^{+0.001}$ & $0.512_{-0.145}^{+0.172}$ & $1.76_{-1.114}^{+0.958}$ & $0.446$ \\
270.01 & $0.057_{-0.001}^{+0.001}$ & $0.208_{-0.142}^{+0.181}$ & $0.07_{-0.001}^{+0.001}$ & $0.147_{-0.093}^{+0.308}$ & $1.672_{-1.843}^{+1.619}$ & $0.056$ \\
\enddata
\tablecomments{We report the median, $16^{th}$ and $84^{th}$ percentile for each parameter. The eccentricity posteriors are rarely described well by a normal distribution, so we report the median, $16^{th}$ and $84^{th}$ percentile along with $e_{mode}$, the statistical mode of the marginalized eccentricity posterior. Only a portion of this table is shown here to demonstrate its form and function. The full table is available online in machine-readable form.}
\end{deluxetable*}

\subsection{Inference of the Underlying Eccentricity Distribution}

Given the set of $K$ individual eccentricity posteriors for $K$ planets and a parent distribution model with parameters $\theta$, we construct a framework by which to infer the underlying, population eccentricity distribution. The combined likelihood function for the underlying population distribution (assuming uniform transit probability) is $p(obs|\theta)$ is

\begin{equation}
    p(obs | \theta) = \prod_{k=1}^{K} \frac{1}{N} \sum_{n=1}^{N} \frac{p{(e_k^n|\theta)}}{p(e_k^n|\alpha)}
\end{equation}

where the number of planets is $K$ and each posterior distribution has $N$ samples. We assume that the $e$ posteriors of all planets are independent and sampled from a uniform prior, so we take $p(\mathbf{e_k^n}|\alpha) =1$. 

The geometric transit probability is non-uniform with eccentricity, and planets with high eccentricities are likelier to transit due to their enhanced transit probability at periapse \citep{barnes_effects_2007, burke_impact_2008, kipping_bayesian_2014}. 

For a planet that transits its star (indicated by a boolean $\hat{t}$), the transit probability is

\begin{equation}
    p(e,\omega|\hat{t}) = \biggl( \frac{1 + e \sin \omega}{1-e^2} \biggr).
\end{equation}
We integrate the reciprocal of the transit probability into the likelihood function to account for the non-uniform transit probability; the marginalized likelihood $p(obs | \theta)$ becomes 

\begin{equation}
    p(obs | \theta) = \prod_{k=1}^{K} \frac{1}{N} \sum_{n=1}^{N} p{(e_k^n|\theta)} \biggl( \frac{1-e{_k^n}^2}{1 + e{_k^n} \sin \omega{_k^n}} \biggr)
\end{equation}

\subsubsection{Model Choice for the Underlying Eccentricity Distribution}

Within a Bayesian hierarchical framework, we infer the underlying eccentricity distribution from the individual $e$ posteriors. Intrinsically, we must select some model distribution that accurately represents the shape of the true, underlying eccentricity distribution.

There is a strong physical motivation for the usage of the Rayleigh distribution to model the underlying orbital eccentricity distribution. If we assume the orbital mid-planes of planetary systems is randomly distributed with respect to our line of sight, and the inclinations of individual planets within a multi-planet system are distributed normally around the orbital mid-plane, it then follows that mutual inclinations between pairs of planets are distributed according to a Rayleigh distribution. Recent studies (e.g. \citealt{he_architectures_2020} and \citealt{millholland_evidence_2021}) show evidence that eccentricities follow mutual inclinations in being distributed according to the system's maximum angular momentum deficit stability limit; in other words, that eccentricities are distributed in a similar pattern to mutual inclinations. These arguments would suggest that orbital eccentricities are also distributed according to a Rayleigh distribution.

However, it is becoming increasingly clear that the Rayleigh distribution does not accurately describe the shape of the underlying eccentricity distribution when measured via the photoeccentric effect (e.g. \citetalias{sagear_orbital_2023} and \citetalias{gilbert_planets_2025}). Population eccentricity posterior distributions tend to follow a distinct shape, usually peaking at $e=0$ and falling off quasi-exponentially to $e=1$. We first elect to describe the underlying eccentricity distribution using a flexible, non-parametric histogram model (following \citetalias{gilbert_planets_2025}); then, for full reproducibility and computational efficiency, we analyze the eccentricity distribution using an analytical model that closely follows the shape of the best-fit flexible histogram model.

The process of applying a flexible, non-parametric histogram model to fit an underlying eccentricity distribution is thoroughly described in \citetalias{gilbert_planets_2025}; for further details on implementing such a non-parametric likelihood function, see \citet{foreman-mackey_exoplanet_2014} and \citet{masuda_inferring_2022}. We construct an empirical histogram model using the eccentricity posteriors of our planetary sample. We include an overview of this process in Appendix \ref{appendix:empirical}.

Ultimately, we lack sufficient physical understanding of the shape of the underlying eccentricity distribution. Though a smoothed and scaled flexible histogram model may fit the combined eccentricity posteriors more precisely, we elect to make a tradeoff between model flexibility and full reproducibility with computational efficiency. Therefore, we select an analytic function to describe the eccentricity model using the best-fit flexible histogram model as a point of comparison. We choose to use a Beta distribution defined in the range $[0, 1]$ with a modified prior which mandates the function always peaks at 0 and monotonically decreases ($\alpha < 1$ and $\beta > 1$). 
Care must be taken when constructing and sampling across Bayesian hierarchical models, especially when modeling Beta distributions; \citet{nagpal_impact_2023} and \citet{do_o_orbital_2023} have shown that improper choices of prior distributions could severely bias results, specifically in cases of modeling orbital eccentricity parent distributions as Beta distributions. To mitigate such biases, we reparametrize with $\tau = 1/\sqrt{\alpha + \beta}$ and $\mu = \alpha/(\alpha + \beta)$. To enforce $\alpha < 1$ and $\beta > 1$, we apply the priors $\tau \sim U(0,1]$ and $\mu \sim U(0,\mathrm{min}(\tau^2, 1-\tau^2)]$. To reduce the effects of strong correlations between parameters (as described in \citealt{betancourt_hamiltonian_2013}), we sample using an off-centered parameterization by directly sampling from $\mu_{\textrm{latent}}$ with the prior $\mu_{\textrm{latent}} \sim U(0,1]$, where $\mu = \mu_{l} + \mu_{\textrm{latent}}(\mu_{l}+\mu_{u})$ and $\mu_{l}$ and $\mu_{u}$ are the lower and upper bounds of the prior on $\mu$, respectively. We present the eccentricity distribution for planets in five radius bins modeled using our flexible histogram model in Appendix \ref{appendix:empirical}. Though we elect to model the eccentricity distribution using the Beta distribution, the process of generating the best-fit empirical model was critical in ensuring that we select a model and priors that match the shape of the empirical eccentricity distribution.

\subsection{Treatment of planet radius uncertainty}

In order to investigate the eccentricity-radius relation, robust and consistent planet radii measurements are needed. As described in Section \ref{subsec:planetsample}, we calculate each stellar host's radius using the empirical \citet{mann_how_2015} relation. With transit fitting, we obtain a planet-to-star radius ratio posterior distribution for each TOI; we multiply the calculated stellar radius by the radius ratio posterior to obtain a distribution estimate of the planet radius. 

We bin the radii, and most planets have radius uncertainties comparable to the bin size (10 percent). Following \citet{newton_approximate_1994} and \citetalias{gilbert_planets_2025}, we adjust the hierarchical Bayesian likelihood function to account for the planet radii uncertainties:
\begin{equation} \label{eq:likelihood_w_radius}
    \mathcal{L}_{\theta} = \prod_{k=1}^K{\biggl [\frac{1}{K}\sum_{n=1}^N{f_{\alpha} (e_{k}^n) \biggl( \frac{1-{e^n_{k}}^2}{1+e_{k}^n \sin \omega_{k}^n} \biggr) } \biggr]^{w_n}}
\end{equation}
where the weights $w_{n}$ are calculated by defining each radius bin as a Gaussian, whose likelihood function is

\begin{equation}
    g(R) = \frac{1}{\Delta R \sqrt{2\pi}} \mathrm{exp} \biggl( \frac{-(R-R_0)^2}{2\Delta R^2} \biggr).
\end{equation}
$R_0$ is the planet's corresponding bin center and $\Delta R$ is the fractional full-width-at-half-maximum corresponding to the bin width divided by the bin center. For each planet, we sum over all $N$ radius ``posterior" samples with

\begin{equation}
    w_n = \frac{1}{N} \sum_{n=1}^N{g(R_{nk})} \mathrm{,}
\end{equation}
normalizing the weights such that the expected value $\langle w \rangle = 1$.

\subsection{Compressing eccentricity distributions into $\langle e \rangle$}

Though the full posterior probability function is a more accurate representation of the eccentricity distribution, for practicality we must choose a compressed value $\langle e \rangle$ to represent the $e$ distribution. For any given sample of $e$ posteriors, we evaluate Equation \ref{eq:likelihood_w_radius} with $f_{\theta}(e^n_k)$ modeled as a Beta distribution with $\alpha < 1$ and $\beta < 1$. We sample the model parameters $\alpha$ and $\beta$ with \texttt{numpyro} \citep{phan_composable_2019} using the NUTS sampler \citep{homan_no-u-turn_2014} for 1000 tuning steps and 1000 posterior draws. This leaves us with a posterior for each of the model parameters. We evaluate all of the posterior beta distributions across the range $[0,1]$. We calculate the mean value of each evaluated model. We take the mean of these median values as \avgenosp, and the 16th and 84th percentiles of these median values as the lower and upper errors on \avgenosp.

\section{Results} \label{sec:results}

With individual eccentricity posteriors in hand for both the \textit{Kepler} and \textit{TESS} M dwarf planet samples, we consider the eccentricity--radius relation for planets orbiting M dwarfs within the hierarchical Bayesian framework described in Section \ref{sec:analysis}. The flexible hierarchical analysis technique enables us to analyze and compare different subsamples of planets along several axes. We consider first how the results vary with transit survey and planet multiplicity before investigating the relationship with radius. 

\subsection{Transit survey and planet multiplicity}\label{subsec:keplertessplots}

The extent to which a given transit survey is sensitive to the architecture of observed planetary systems is a function of the survey itself (that is, the photometric precision and duration of observations) and also of the geometric limitations of the transit detection method. A transit survey will inherently have lower completeness (that is, detect fewer transiting planets in a multi-planet system around a given star) for systems with higher mutual inclination. Planetary system properties such as number of planets per star, average orbital eccentricity, average mutual inclination, and average spacing between planets vary together \citep{fabrycky_architecture_2014, xie_exoplanet_2016, ballard_kepler_2016, van_eylen_orbital_2019} and also shape the resulting yield of any transit survey. For this reason, an examination of the eccentricities of ``multi-transit" versus ``single-transit" requires care, and doubly so when comparing the results of two transit surveys. As described in Section 1, eccentricities for \textit{Kepler} exoplanets are higher in the single-transit population as compared to the multi-transit population \citep{xie_exoplanet_2016, van_eylen_orbital_2019, sagear_orbital_2023}. This result is consistent with an underlying range of dynamical temperatures among planetary systems \citep{tremaine_statistical_2015}, for which single-transit systems sample the ``hottest" end. \citetalias{sagear_orbital_2023} found evidence that the single-transit eccentricity population is actually best described as two populations, consistent with a scenario in which single-transits are a mixture: some drawn from authentically dynamically hotter (that is, higher eccentricity) systems and some drawn from dynamically colder (that is, lower eccentricity) systems for which the viewer is geometrically unlucky. Previous work has quantified these fractions for \textit{Kepler} versus \textit{TESS} M dwarfs: while between $15$-$20\%$ of M dwarfs host densely packed multi-planet systems with low orbital eccentricity \citep{muirhead_characterizing_2014, ballard_kepler_2016}, they are overrepresented in transit surveys. For \textit{Kepler}, they likely comprise $50\%$ of detected systems, and for \textit{TESS} the overrepresentation is even stronger: such systems likely make up $\sim 75\%$ of the small-star planet yield \citep{ballard_predicted_2019}. 

\begin{figure}
    \centering
    \includegraphics[width=\linewidth]{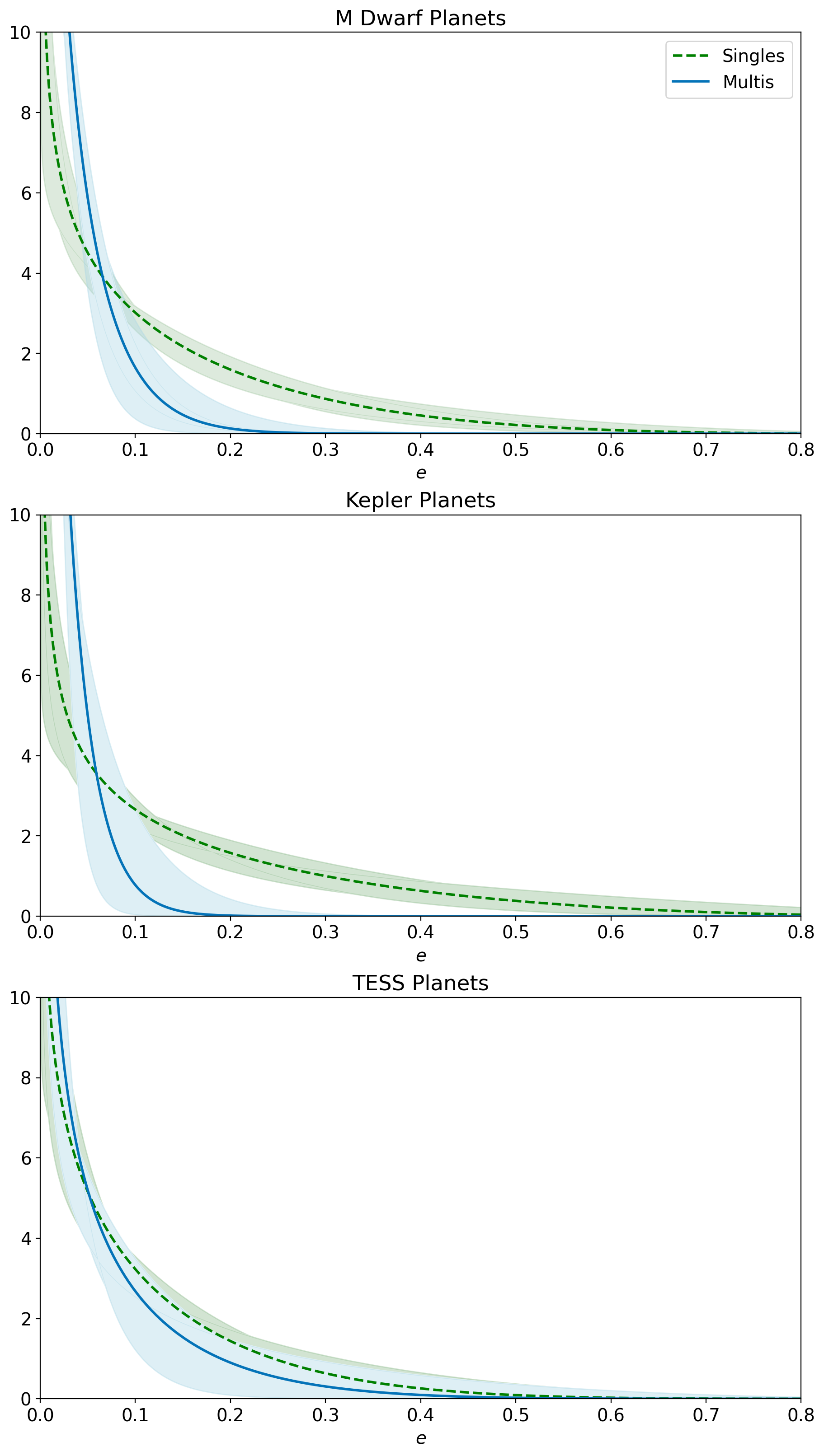}
    \caption{\textit{Top:} Best-fit parent eccentricity distributions for the entire sample of singles 
    (green) and multis (blue) in this work. \textit{Middle:} Best-fit parent eccentricity distributions for singles and multis for the sample of M dwarf KOIs in this work. The best-fit distributions for both populations are distinct, as shown in \citetalias{sagear_orbital_2023}. \textit{Bottom:} Same as left panel, but for the sample of TOIs in this work. The difference in parent distributions for singles and multis is less apparent for TOIs, likely due to factors associated with the \textit{TESS} observation strategy leading to higher levels of ``contamination" in the singles population by true multi-planet systems.}
    \label{fig:kepler_tess}
\end{figure}

With this context in hand, we briefly compare orbital eccentricities for singles and multis discovered by \textit{Kepler} and \textit{TESS}, and discuss differences in the eccentricity distributions between surveys. In the top panel of Figure \ref{fig:kepler_tess}, we show the best-fit Beta distribution for all singles and multis in our sample. In the middle panel of Figure \ref{fig:kepler_tess}, we show the best-fit Beta distribution for \textit{Kepler} M dwarf singles and multis, and the same plot for \textit{TESS} M dwarf singles and multis on the bottom panel. We observe that for \textit{Kepler} M dwarf planets, eccentricities for singles and multis are drawn from distinct parent distributions \citetalias{sagear_orbital_2023}, with singles having an elevated eccentricity. This is not obviously the case for 
\textit{TESS} planets: instead, eccentricities for singles and multis appear consistent. This difference may be explained by the demographic differences in small planets discovered by \textit{Kepler} vs. \textit{TESS}.  \textit{Kepler}'s years-long baseline observation time \citep{borucki_kepler_2010} means that multi-planet systems are more likely to be identified as such, especially where one member's orbital period is longer than tens of days. Meanwhile, \textit{TESS} observes stars with a much shorter baseline observation time than \textit{Kepler} targets, even for stars within the Continuous Viewing Zone \citep{stassun_tess_2018}. For \textit{TESS}-discovered single-transit planets, it is more likely that another member of the planetary system lurks undiscovered, especially if that hidden planet has a period longer than one third of the baseline observation time (only $\sim 10$ days) \citep{ballard_predicted_2019}. This is likely because the two surveys are sampling different components of the underlying distribution of dynamical temperature among planetary systems. While the \textit{Kepler} single-transit dynamically hotter sample is ``contaminated" to some degree with dynamically cold systems (statistically demonstrated in \citealt{ballard_kepler_2016} and \citetalias{sagear_orbital_2023} to be $\sim50\%$), this effect is predicted to be even greater for the \textit{TESS} yield. This is because the mission observed given stars for shorter durations of time and with lower photometric precision, so that even authentic multi-transit systems will not be detected as such by \textit{TESS} \citep{ballard_predicted_2019}. This difference in the degree of single-transit sample contamination is consistent with the overall dynamical coolness of \textit{TESS} systems among both ``singles" and ``multis." 

\subsection{The Eccentricity\textemdash Radius Relationship for Planets Orbiting M Dwarfs}\label{subsec:eccradius}

\begin{figure}[t!]
    \centering
    \includegraphics[width=\linewidth]{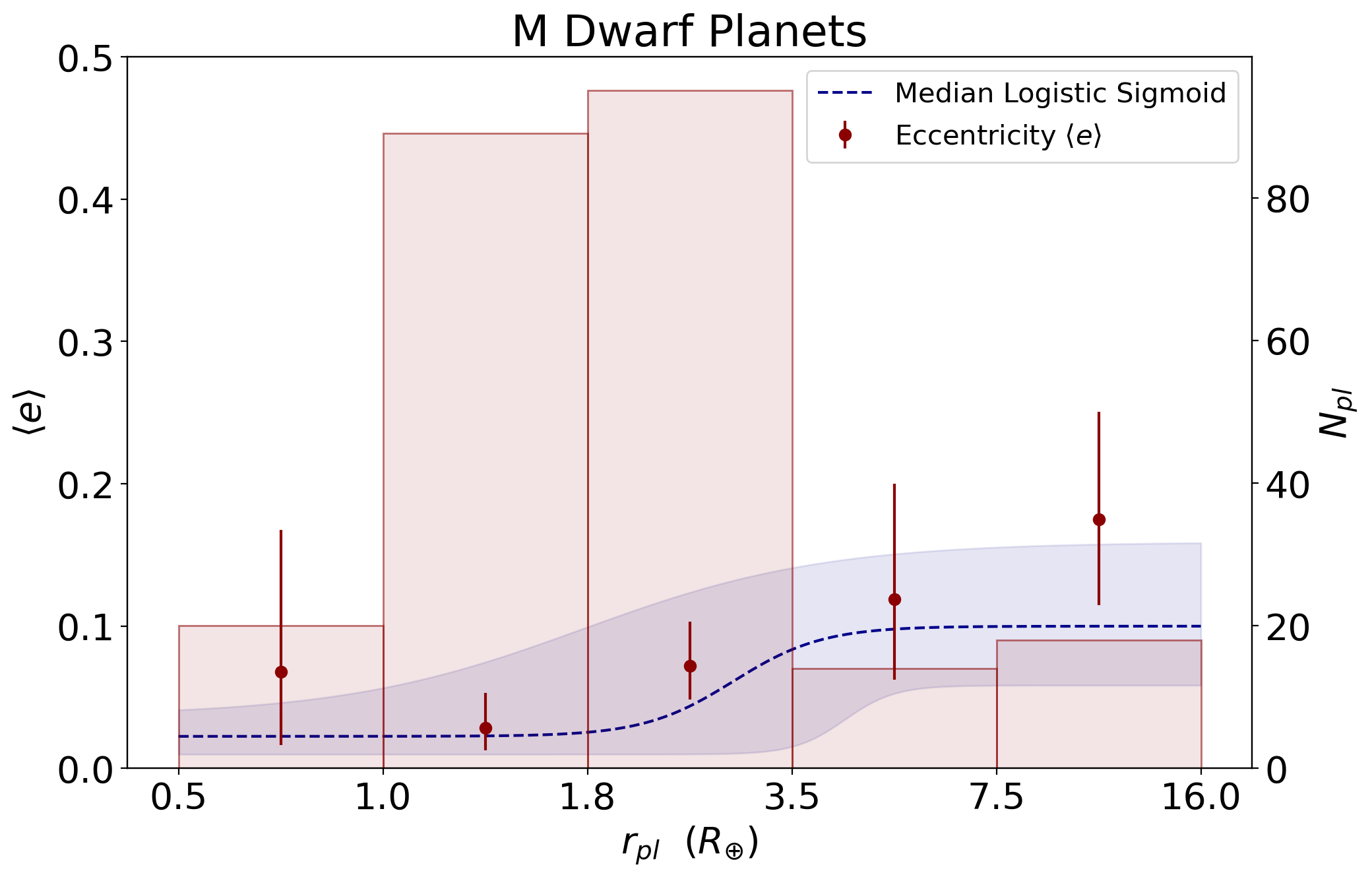}
    \caption{$\langle e \rangle$ vs. planet radius bins for the sample of M dwarf planets (red points). The red error bars present the $16^{th}$ and $84^{th}$ percentiles of $\langle e \rangle$. The red histogram represents the number of M dwarf planets in each radius bin (right-hand vertical axis). The median logistic sigmoid model is plotted in dark blue. The $16^{th}$ and $84^{th}$ percentiles of the logistic sigmoid are denoted by the dark blue shaded regions.}
    \label{fig:eccradius_linear}
\end{figure}

We explore the relationship between eccentricity and binned planet radius. We bin the sample of planets based on radius according to the bins in Table \ref{tab:rad_bin_edges}. In Figure \ref{fig:eccradius_linear}, we present the eccentricity\textemdash radius relation divided among these bins for the entire sample of planets, along with the sample size for each bin. We see evidence for elevated eccentricities for planets with $r > 3.5 R_{\oplus}$. To quantify the transition, as in \citetalias{gilbert_planets_2025}, we fit a logistic sigmoid to the curve with

\begin{equation} \label{eq:sigmoid}
    f(x) = B + \frac{L}{1+e^{-k(x-x_t)}}. 
\end{equation}
where $B$ represents the baseline eccentricity, $L$ represents the sigmoid normalization value, and $x_t$ and $k$ represent the location and rate of change of the transition from low to high $e$, respectively. The median sigmoid along with its 16th and 84th percentile errors are plotted in Figure \ref{fig:eccradius_linear}. The transition from low to high eccentricity occurs at $3.1_{- 1.2}^{+1.5}$ $R_{\oplus}$. \citetalias{gilbert_planets_2025} found that the transition for singles orbiting FGK dwarfs occurs at $3.3 \pm 0.4$ $R_{\oplus}$ and for multis at $4.2 \pm 0.9$ $R_{\oplus}$, making the transition radii from low to high eccentricities consistent within $1\sigma$ for FGK dwarf singles, FGK dwarf multis, and M dwarf planets. In terms of the logistic sigmoid functional form (Equation \ref{eq:sigmoid}), the low eccentricity $e_{low} = B$ and the high eccentricity $e_{high} = B+L$. For M dwarf planets, $e_{\mathrm{high}}/e_{\mathrm{low}} = 4.6_{-1.9}^{+5.8}$, which is also within $1\sigma$ of the findings of \citetalias{gilbert_planets_2025}.

\begin{deluxetable*}{cccccc}
\tablecaption{Radius bin locations and associated planets \label{tab:rad_bin_edges}}
\tablehead{\colhead{Bin edges ($R_{\oplus}$)} & \colhead{$N_{pl}$} & \colhead{$N_{pl}$ (\textit{Kepler})} & \colhead{$N_{pl}$ (\textit{TESS})} & \colhead{$N_{pl}$ (Singles)} & \colhead{$N_{pl}$ (Multis)}}
\startdata
0.5\textemdash1.0 & 20 & 19 & 1 & 7 & 13\\
1.0\textemdash1.8 & 89 & 62 & 27 & 35 & 54\\
1.8\textemdash3.5 & 95 & 66 & 29 & 37 & 58\\
3.5\textemdash7.5 & 14 & 7 & 7 & 10 & 4\\
7.5\textemdash16.0 & 18 & 3 & 15 & 18 & 0\\
\enddata
\end{deluxetable*}

Rather than being defined by a single planet radius, the radius valley shifts to a larger radius for shorter orbital periods \citep{fulton_california-kepler_2017, fulton_california-kepler_2018, vaneylen_asteroseismic_2018, ho_shallower_2024}. For this reason, ``radius gap" planets reside along a diagonal in period-radius parameter space, and so we construct our bins to be split along diagonals in period\textemdash radius space. We construct our bins based on the relation derived by support-vector machine (SVM) from \citet{ho_shallower_2024}: 

\begin{multline}
 \label{eq:ho_24}
    \log_{10}(R_p/R_{\oplus}) = -0.09 \log_{10}(P/d) \\
    + 0.21 \log_{10} (M_{\star}/M_{\odot}) + C, 
\end{multline}
where $M_{\star}$ is the median stellar mass of the sample. Using stellar masses derived with the empirical relation from \citet{mann_how_2019}, we evaluate this model using our median stellar mass of $0.55 R_{\odot}$. We define our bins by varying $C$ in Equation \ref{eq:ho_24} between $0.0$ and $1.0$ with width $0.095$, resulting in parallel slices in $log(P)-log(r_{p})$ space (see Figure 4 of \citet{ho_shallower_2024} for an approximate visual representation). We split our sample along these period\textemdash radius bins and follow the Bayesian hierarchical modeling procedure described in Section \ref{sec:analysis}. Therefore, the ``radius" of each bin here refers to $r_{\textrm{adj}}$, which we define as the center of the period\textemdash radius diagonal slice evaluated at $P=10$ days.

\begin{figure*}[t!]
    \centering
    \includegraphics[width=\linewidth]{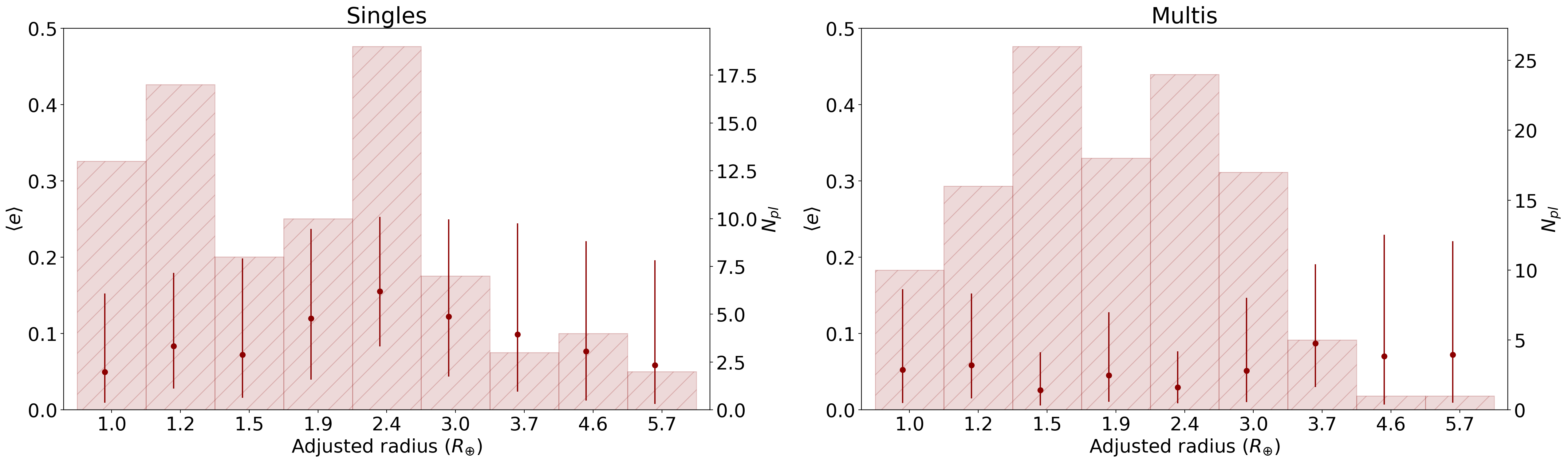}
    \caption{\textit{Left:} $\langle e \rangle$ vs. adjusted planet radius $r_{\textrm{adj}}$ bin centers for singles with $0.5 < r_{\textrm{adj}} < 5.1$ $R_{\oplus}$ (red points). The red error bars present the $16^{th}$ and $84^{th}$ percentiles of $\langle e \rangle$. The red histogram represents the number of M dwarf planets in each bin (right-hand vertical axis).} \textit{Right:} Same as left panel, but for multis.
    \label{fig:eccradius_diag}
\end{figure*}

In Figure \ref{fig:eccradius_diag}, we show how this radius-eccentricity relationship varies between the sample of single-transit and multi-transit systems. While single-transit systems have higher eccentricity overall, it does not present in a completely uniform way across radius. Rather, singles between adjusted radii $1.9$ and $3.0$ $R_{\oplus}$ exhibit modestly elevated eccentricities, though with only modest statistical significance. Meanwhile, the population of multis appears to maintain very low eccentricities across all radii. 

We quantify the significance of the potential eccentricity peak in the singles population between adjusted radii $1.9$ and $3.0$ $R_{\oplus}$. We fit a simple Gaussian with a variable baseline to the data using

\begin{equation}
    f(x) = B + Ae^{-0.5(x-x_p)^2/s^2},
\end{equation}
where $B$ denotes the baseline eccentricity, $A$ denotes the Gaussian peak amplitude, $x_p$ denotes the peak location, and $s$ denotes the peak width. Setting $A=0$ in this case recovers a flat line across radius, and the extent to which $A$ is inconsistent with zero indicates the deviation of the data from the flat line scenario.   As in \citetalias{gilbert_planets_2025}, we allow $A$ to float between positive and negative values, and allow the proportion of positive $A$ values to inform the significance of eccentricity peak. 

For the singles, $A > 0$ for $69 \%$ of posterior samples, corresponding to a deviation from a flat line at the roughly $1\sigma$ level of confidence: that is, the data are consistent with a flat line. For the multis, there is no preference whatsoever between positive and negative $A$ values in the posterior samples ($A > 0$ for $49 \%$ of posterior samples). While a $1\sigma$ level of significance is insufficient to claim detection or non-detection of an eccentricity peak, this result is unsurprising given that \citetalias{gilbert_planets_2025} detected the peak to only $2.1\sigma$ with a planetary sample over six times larger than the sample in this work. We discuss these results in detail, along with possible physical interpretations and comparisons between planet populations around various stellar hosts, in Section \ref{sec:discussion}.

\section{Discussion} \label{sec:discussion}

\subsection{The Eccentricity\textemdash Radius Relationship for Planets Across Spectral types: M dwarfs vs FGK dwarfs} 
\label{subsec:gilbertcomparison}

We have crafted this experiment to test a specific question: does the eccentricity-radius relationship for M dwarfs resemble that of FGK dwarfs? To this end, we framed this study to enable direct comparison to \citetalias{gilbert_planets_2025}, who measured eccentricities for 1,646 planets orbiting FGK stars via their \textit{Kepler} light curves. \citetalias{gilbert_planets_2025} find that planets larger than $3.5$ $R_{\oplus}$ transition sharply to relatively elevated eccentricities, and planets in the radius gap (between roughly $1.5$ and $2$ $R_{\oplus}$) have modestly higher eccentricities than those just outside the radius gap. 

\begin{figure}
    \centering
    \includegraphics[width=\linewidth]{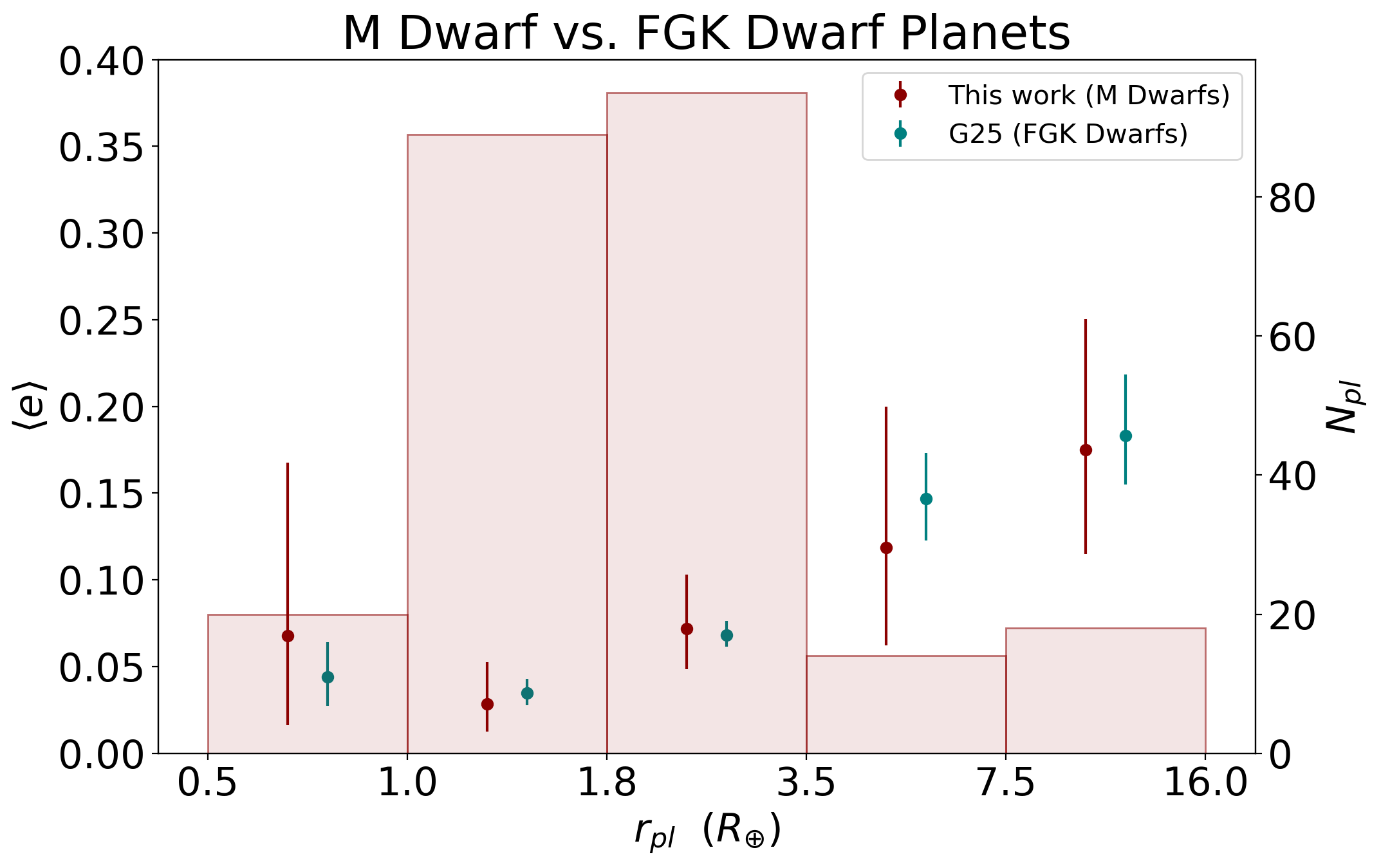}
    \caption{\avge vs. planet radius bin centers for the sample of M dwarf planets in this work (red points) and for FGK dwarf planets from \citetalias{gilbert_planets_2025} (teal points). The error bars represent the $16^{th}$ and $84^{th}$ percentiles of $\langle e \rangle$. The red histogram represents the number of M dwarf planets in each radius bin (right-hand vertical axis).}
    \label{fig:binnedgilbertcoarse}
\end{figure}

\begin{figure*}[t!]
    \centering
    \includegraphics[width=\linewidth]{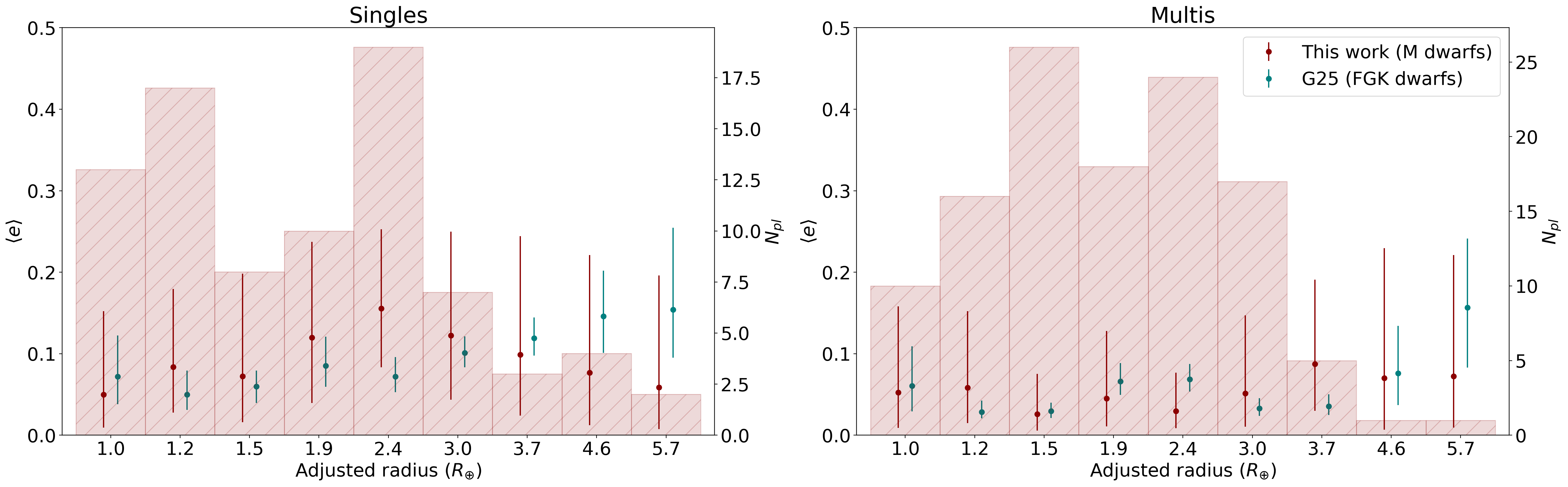}
    \caption{\textit{Left:} $\langle e \rangle$ vs. adjusted planet radius bin centers for the sample of M dwarf singles in this work (red points) and for FGK dwarf singles from \citetalias{gilbert_planets_2025} (teal points), according to the diagonal binning mechanism described in Section \ref{subsec:gilbertcomparison}. The error bars represent the $16^{th}$ and $84^{th}$ percentiles of $\langle e \rangle$. The red histogram represents the number of M dwarf planets in each radius bin (right-hand vertical axis). \textit{Right:} Same as left panel, but for multis.}
    \label{fig:binnedgilbertdiagonal}
\end{figure*}

Because the sample of available M dwarf planet eccentricity posteriors across \citetalias{sagear_orbital_2023} and this work is much smaller than that of \citetalias{gilbert_planets_2025} (\nplanets vs. 1,646), we bin the \citetalias{gilbert_planets_2025} eccentricity\textemdash radius sample to match the larger bins in this work. To enable a one-to-one comparison, we employ the 1,646 eccentricity posteriors from that work to exactly replicate the analysis in Section \ref{subsec:eccradius}: we again assume a Beta distribution for the underlying parent distribution, enforcing the priors $\alpha < 1$ and $\beta > 1$, and use the exact same binning scheme.  In Figure \ref{fig:binnedgilbertcoarse}, we compare the newly binned \citetalias{gilbert_planets_2025} relation and the binned M dwarf eccentricity\textemdash radius relation from this work. In Figure \ref{fig:binnedgilbertdiagonal}, we compare the \citetalias{gilbert_planets_2025} relation with the M dwarf relation according to the diagonal bins. The \avge results from this work (M dwarf planets) and \citetalias{gilbert_planets_2025} (FGK dwarf planets) are generally consistent (insofar as \avge for the two samples overlap significantly in each diagonal radius bin, and entirely in each coarse radius bin).

A main point of interest in comparing the eccentricities of M dwarf and FGK dwarf planets is the prevalence of elevated eccentricities at the radius gap. \citetalias{gilbert_planets_2025} found modest evidence for elevated eccentricities for radius gap planets orbiting FGK dwarfs for both singles and multis. As discussed in Section \ref{sec:results}, across radius bins, a model with a lower baseline and elevated eccentricities between $1.9$ and $3.0$ $R_{\oplus}$ for M dwarf singles is indistinguishable from a flat line at a slightly higher baseline (roughly $1\sigma$). For multis, there is also a lack of significant evidence for elevated eccentricities at any radius, including near the radius gap. 

While more precise \avge measurements are needed to confirm this disagreement (either in the form of a larger planetary sample or more precise individual $e$ measurements), if this disagreement is indeed physical in nature, it may result from a difference in the mutual Hill spacing of exoplanets orbiting M dwarf and FGK dwarf systems. \cite{mulders_slar-mass-dependent_2015} found that M dwarf exoplanet systems, between orbital periods of 2-50 days, contain approximately twice the planet mass of systems orbiting G dwarfs in the same period range. \citet{moriarty_kepler_2016}, via $N$-body simulations of late-stage, in-situ planet formation, similarly found that M dwarf systems must be significantly more closely packed to replicate the \textit{Kepler} yield (that is, the typical mutual Hill radius between planets orbiting M dwarfs is smaller than the corresponding Hill radius for planets orbiting FGK dwarfs) \citep{hamilton_orbital_1992}. Whatever the mechanism responsible for elevated eccentricities in radius gap planets (planet-planet collisions, for example) \citep{ho_deep_2023}, this difference in mutual Hill spacing distributions may allow FGK dwarf planets in the radius gap to maintain slightly higher eccentricities while still maintaining stability in less tightly-packed multi-planet systems than M dwarfs typically host.

\subsection{Effects of Stellar Metallicity on Orbital Eccentricity} \label{subsec:metallicity}

We investigate the role of host stellar metallicity on the eccentricity trends we observe. In the case where high orbital eccentricities are generated through gravitational planet-planet interactions, especially among large planets, we may expect to see a positive correlation between eccentricity and stellar metallicity \citep{dawson_giant_2013}, especially among the largest planets \citepalias{gilbert_planets_2025}. We seek to determine whether the single-transit radius bins that exhibit elevated eccentricities (with adjusted radii between $1.9$ and $3.0$ $R_{\oplus}$) might be directly correlated with elevated host stellar metallicities.

\begin{deluxetable*}{ccccc}
\tablecaption{Spectroscopic metallicities and provenances for M dwarf TOIs. \label{tab:metallicityprovenance}}
\tablehead{\colhead{TIC} & \colhead{[Fe/H]} & \colhead{$\sigma_{[Fe/H], l}$} & \colhead{$\sigma_{[Fe/H], u}$} & \colhead{Provenance}}
\startdata
12421862 & -0.36 & -0.09 & 0.09 & \citet{kuznetsov_characterization_2019} \\
29960110 & 0.11 & -0.19 & 0.19 & \citet{rains_characterization_2021} \\
44737596 & 0.21 & -0.19 & 0.19 & \citet{rains_characterization_2021} \\
70899085 & 0.09 & -0.19 & 0.19 & \citet{rains_characterization_2021} \\
77156829 & -0.45 & -0.19 & 0.19 & \citet{rains_characterization_2021} \\
86263325 & 0.04 & -0.12 & 0.12 & \citet{libby-roberts_-depth_2023} \\
98796344 & -0.27 & -0.19 & 0.19 & \citet{rains_characterization_2021} \\
103633434 & -0.33 & -0.16 & 0.16 & \citet{bluhm_precise_2020} \\
118327550 & -0.08 & -0.13 & 0.13 & \citet{gore_metallicities_2024} \\
119584412 & 0.15 & -0.04 & 0.05 & \citet{macdougall_tess-keck_2023} \\
\enddata
\tablecomments{The entire list of stellar provenances is as follows: \citet{abdurrouf_seventeenth_2022, almenara_toi-4860_2024, antoniadis-karnavas_odusseas_2024, bakos_hats-71b_2020, bluhm_precise_2020, buder_galah_2021, costa-almeida_m_2021, gan_metallicity_2025, ghachoui_tess_2023, gore_metallicities_2024, hirano_earth-sized_2023, jonsson_apogee_2020, kuznetsov_characterization_2019, libby-roberts_-depth_2023, macdougall_accurate_2023, marfil_carmenes_2021, rains_characterization_2021, timmermans_toi-4336_2024}. Only a portion of this table is shown here to demonstrate its form and function. The full table is available online in machine-readable form.}
\end{deluxetable*}

For KOI hosts, we use spectroscopic metallicities from \citet{muirhead_characterizing_2014}. For TOI hosts, we conduct a literature search for spectroscopically-determined stellar metallicities. We present the metallicities and their sources in Table \ref{tab:metallicityprovenance}. For each diagonal period\textemdash radius bin, we take the associated stellar host metallicities $\mu_i$ and their individual measurement uncertainties $\sigma_i$ and calculate the mean and propagated errors, assuming Gaussian uncertainties.

\begin{figure*}
    \centering
    \includegraphics[width=\linewidth]{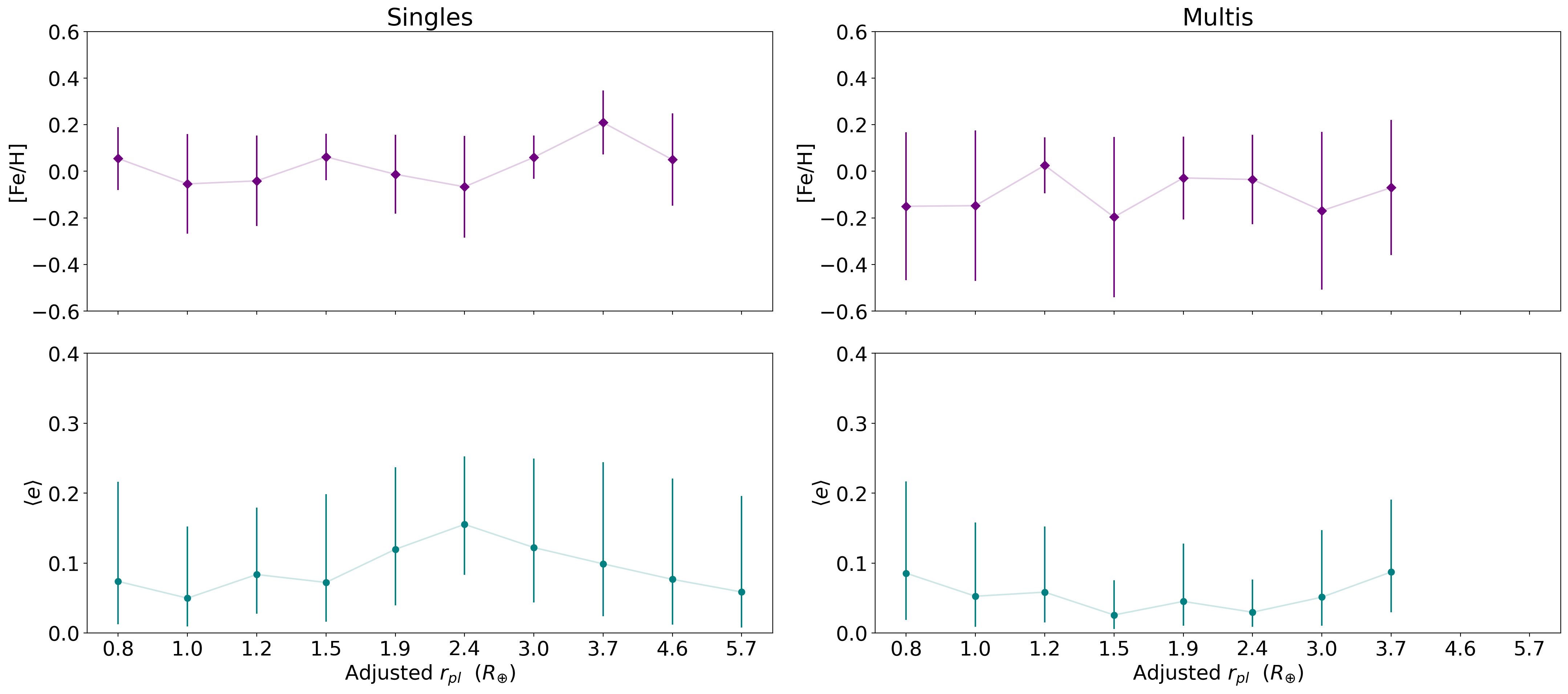}
    \caption{\textit{Left:} On the top panel, $[Fe/H]$ for single-transit planet hosts associated with each adjusted radius bin (purple). On the bottom panel, $\langle e \rangle$ vs. adjusted planet radius bin centers for the sample of M dwarf singles in this work (teal). \textit{Right:} Same as left panel, but for multis.}
    \label{fig:metallicity}
\end{figure*}

In Figure \ref{fig:metallicity}, we show stellar metallicities for the host stars associated with each radius bin on the top panels, and \avge in each $R_p$ bin (as in Figures \ref{fig:eccradius_diag} and \ref{fig:binnedgilbertdiagonal}) on the bottom panels. Multi hosts tend to be more metal-poor on average than hosts of singles \citep{brewer_compact_2018, anderson_higher_2021}, a trend we see reflected in our stellar sample. Notably, the elevated eccentricities we see in singles with radii between $1.9$ and $3.0$ $R_{\oplus}$ do not appear to be associated with significantly higher stellar host metallicities (Figure \ref{fig:metallicity}), suggesting that if eccentricity peak exists at these radii, the peak is unlikely to be fully explained by stellar metallicity, reflecting the findings of \citetalias{gilbert_planets_2025} for small ($< 3.5$ $R_{\oplus}$) planets orbiting FGK dwarfs.

\begin{figure*}
    \centering
    \includegraphics[width=\linewidth]{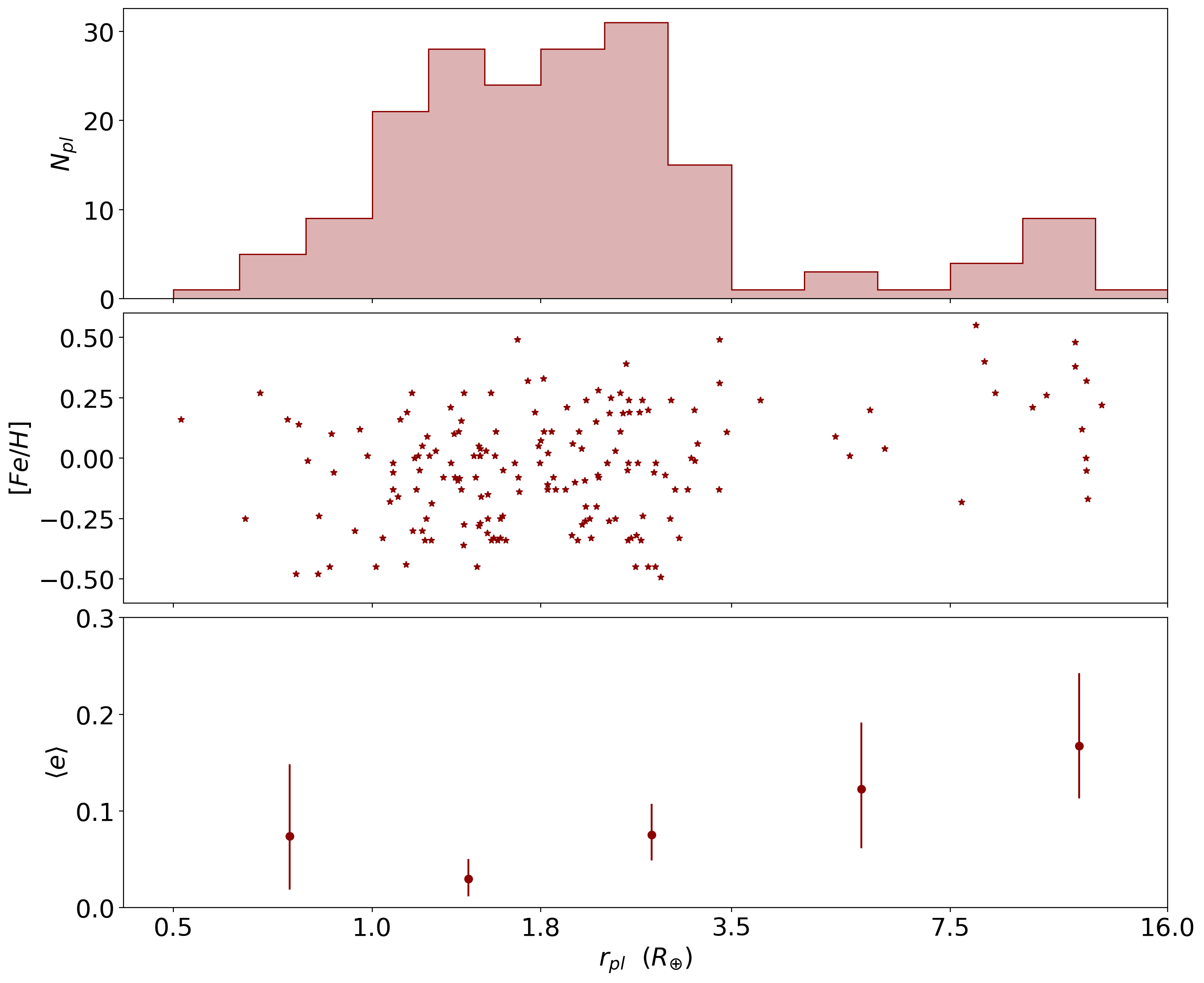}
    \caption{\textit{Top:} Radius distribution of M dwarf planets in this sample. \textit{Middle:} [Fe/H] vs. planet radius for the M dwarf planet sample. \textit{Bottom:} \avge as a function of binned planet radius. As \citetalias{gilbert_planets_2025} showed for planets around FGK stars, we demonstrate an apparent transition in planet occurrence, metallicity, and eccentricity in M dwarf planets.}
    \label{fig:metallicity_panel}
\end{figure*}

We demonstrate that planet occurrence rate, host stellar metallicity, and eccentricity appear to transition around $3.5 R_{\oplus}$ with Figure \ref{fig:metallicity_panel}. Small planets ($< 3.5 R_{\oplus}$) are abundant and tend to occur around metal-rich and metal-poor stars alike, while larger planets are rarer and preferentially occur around metal-rich stars. These trends have been thoroughly discussed in the literature for planets orbiting stars of various spectral types \citep[e.g.,][]{fischer_planet-metallicity_2005, buchhave_abundance_2012}. \citetalias{gilbert_planets_2025} demonstrated that this transition between small and large planets extends to orbital eccentricity as well, suggesting the prevalence of distinct formation pathways for these planet radius categories. With a sample of only \nplanets planets, we show that the relationship between planet radius and occurrence, metallicity, and orbital eccentricity presented in \citetalias{gilbert_planets_2025} extends to planets around M dwarfs, suggesting that the formation and evolutionary mechanisms responsible for this effect likely extend continuously from Solar-mass hosts to the very smallest stars.

\subsection{Implications for atmospheric loss mechanisms} \label{subsec:implications}

We briefly discuss possible implications of our results on the predominance of various atmospheric loss mechanisms, though more work is needed to determine the significance of any one mechanism. M dwarf planets, especially those in multi-transit systems, in the radius gap appear to maintain the same low eccentricities as small planets just outside the radius gap. If this trend is physical in nature, this may suggest that photoevaporation or core-powered mass loss, rather than giant impacts, are the primary mechanisms through which atmospheric loss occurs for M dwarf planets. Simulations have shown that where giant planet impacts dominate atmospheric loss for small planets, planets in the radius gap exhibit markedly elevated eccentricities \citep{chance_evidence_2024}. This theory is further corroborated by \citetalias{gilbert_planets_2025}, an empirical measurement of elevated eccentricities in the radius gap for a planet population likely to be predisposed to giant impacts. For the M dwarf planet population, dominated by small planets and including few Neptune-sized-or-larger planets, giant impacts as a primary atmospheric loss mechanism is less likely due to the simple lack of giant planets. The fact that we see consistently low eccentricities in the multis population as a function of planet radius (where giant planets are even rarer), and possible elevated eccentricities near the radius valley for the singles population, suggests that giant impacts or planet-planet scattering may play roles in both atmospheric loss and dynamical disruption for single-transit systems. However, photoevaporation (an atmospheric loss mechanism leading to minimal dynamical disruption) may dominate for M dwarf planets. Evidence that the radius valley for M dwarf planets may be narrower and shallower than the radius valley for FGK dwarf planets \citep{owen_mapping_2024, ho_shallower_2024} further supports the idea of an M dwarf planet radius valley sculpted by stellar-luminosity-driven mass loss (though photoevaporation alone cannot fully describe the shape of the low-mass host radius valley) \citep{ho_shallower_2024}.

\section{Conclusions} \label{sec:conclusions}

We summarize our results and work as follows:

\begin{itemize}
\item We employ the photoeccentric effect on a sample of \nplanets transiting planets orbiting M dwarfs, gathered from both NASA's \textit{Kepler} and \textit{TESS} missions. We employ the \citet{macdougall_accurate_2023} importance sampling method to extract these individual eccentricity posteriors. We then derive the likeliest underlying parent distribution from which these measurements are drawn, assuming it resembles a Beta distribution. We slice the sample into radius bins to test whether this parent distribution appears to vary as a function of radius. 
\item We show marked evidence for a transition between low and high orbital eccentricities at a radius of $~3.5$ $R_{\oplus}$ for M dwarf planets, reflecting the results of \citetalias{gilbert_planets_2025} for FGK dwarf planets. We present the transition from high to low planet occurrence, varied to high stellar metallicity, and low to high orbital eccentricities all occurring around $~3.5$ $R_{\oplus}$. These trends are suggestive of distinct formation channels for planets $<3.5$ $R_{\oplus}$ and $>3.5$ $R_{\oplus}$, continuous across stellar spectral types from F to M.
\item When considering only planets smaller than 3.5 $R_{\oplus}$, we find no evidence for elevated eccentricities near to the radius gap: in both the popular and single- and multi-transiting systems, the trend with radius is consistent with a flat line within our measurement uncertainty.  We compare these results to those of \citetalias{gilbert_planets_2025} and, while emphasizing the low significance of this detection, we briefly discuss potential implications for atmospheric loss mechanisms if these trends are physical in nature.
\end{itemize}

The abundance of precise transit light curves from the \textit{Kepler} and \textit{TESS} missions coupled with precise stellar density information for virtually all planet hosts enabled by the Gaia mission allow us to constrain eccentricities for larger and larger samples of planets and makes statistical analysis on a demographic scale possible. Understanding the differences in the eccentricity\textemdash radius relation for planets around FGK stars and M dwarfs enables us to understand differences in planet formation processes, atmospheric loss mechanisms, and evolution for planets around different host stars. These demographic studies, coupled with statistical and N-body simulations of system evolution, will create a comprehensive understanding of planet formation and evolution around the predominant habitable planet hosts, from the smallest and most abundant stars to those resembling our own Sun.

\begin{acknowledgments}
We thank Quadry Chance, Natalia Guerrero, Romy Rodriguez, Nazar Budaiev, and Aleksandra Kuznetsova for helpful discussions about this work.

This work has made use of data from the European Space Agency (ESA) mission {\it Gaia} (\url{https://www.cosmos.esa.int/gaia}), processed by the {\it Gaia} Data Processing and Analysis Consortium (DPAC, \url{https://www.cosmos.esa.int/web/gaia/dpac/consortium}). Funding for the DPAC has been provided by national institutions, in particular the institutions participating in the {\it Gaia} Multilateral Agreement. 
This paper includes data collected by the \textit{Kepler} and \textit{TESS} missions and obtained from the MAST data archive at the Space Telescope Science Institute (STScI). The specific observations analyzed can be accessed via \dataset[https://doi.org/10.17909/T9059R]{https://doi.org/10.17909/T9059R} (KIC) and \dataset[https://doi.org/10.17909/fwdt-2x66]{https://doi.org/10.17909/fwdt-2x66} (TIC). STScI is operated by the Association of Universities for Research in Astronomy, Inc., under NASA contract NAS5–26555. Support to MAST for these data is provided by the NASA Office of Space Science via grant NAG5–7584 and by other grants and contracts. This research has made use of the NASA Exoplanet Archive \citep{nasa_exoplanet_archive_kepler_2019}, which is operated by the California Institute of Technology, under contract with the National Aeronautics and Space Administration under the Exoplanet Exploration Program.
\end{acknowledgments}

\facilities{Gaia, Kepler, TESS, Exoplanet Archive, ExoFOP}

\software{numpy \citep{harris_array_2020}, matplotlib \citep{caswell_matplotlibmatplotlib_2024}, astropy \citep{robitaille_astropy_2013, collaboration_astropy_2018, collaboration_astropy_2022}, numpyro \citep{bingham_pyro_2019,phan_composable_2019}}

\appendix

\section{Validation} \label{appendix:validation}

To ensure that the \textit{Kepler} planet eccentricity measurements of \citetalias{sagear_orbital_2023} are directly comparable to the \textit{TESS} planet eccentricity measurements in this work, we analyze a few selected test cases and compare results.

\subsection{\textit{Kepler} Light Curve Fitting with \texttt{exoplanet} and \texttt{jaxoplanet}}

\citetalias{sagear_orbital_2023} used slightly different machinery to fit the transits and generate the eccentricity posteriors. Due to the availability of more efficient transit-fitting techniques, we adjust the transit-fitting methods for this work to improve efficiency and resource usage. Instead of using the Python transit-fitting package \texttt{exoplanet} \citep{foreman-mackey_exoplanet_2021}, we primarily use the transit-fitting package \texttt{jaxoplanet} \citep{hattori_exoplanet-devjaxoplanet_2024} because of its improved efficiency. As we discuss in Section \ref{subsubsec:importance}, we also utilize the \citet{macdougall_accurate_2023} importance sampling method rather than sampling the eccentricity and longitude of periastron parameters directly, as in \citetalias{sagear_orbital_2023}. Because we wish to directly compare the eccentricity posteriors from \citetalias{sagear_orbital_2023} and this work, we must ensure that these two fitting methods are comparable. To do this, we fit a \textit{Kepler} M dwarf planet light curve using both methods and compare the resulting eccentricity posteriors to ensure they are consistent.

\begin{figure*}\label{fig:koi247_exojax}
  \centering
  \includegraphics[width=\linewidth]{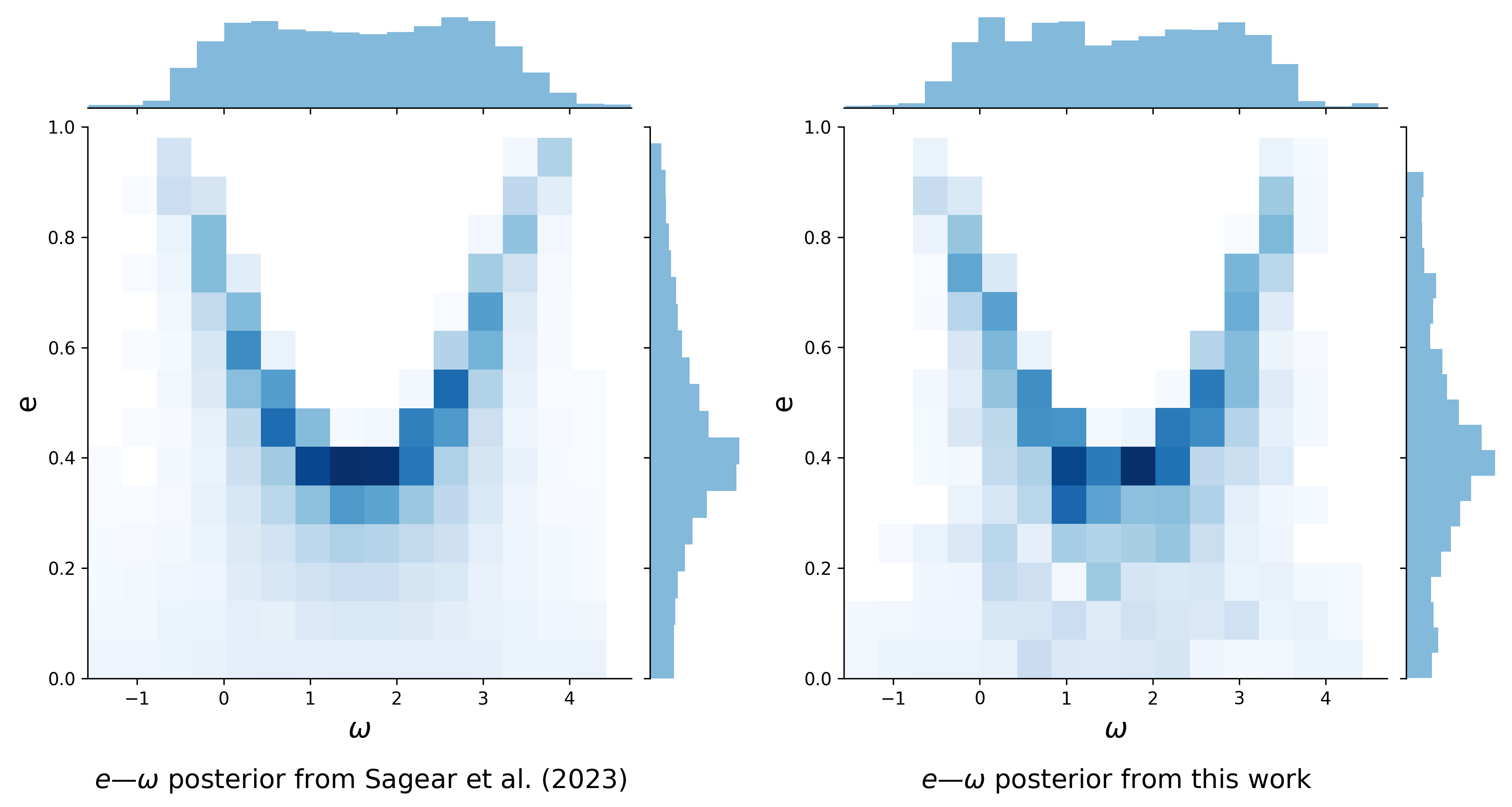}
  \caption{\textit{Left:} $e$ vs. $\omega$ posteriors for KOI 247.01 performed with the method from \citetalias{sagear_orbital_2023}; namely, by directly sampling $\sqrt{e} sin \omega$ and $\sqrt{e} cos \omega$ during the transit fit with the \texttt{exoplanet} software. \textit{Right:} $e$ vs. $\omega$ for KOI 247.01 performed with the method described in Section \ref{sec:analysis}, using the importance sampling method with the \texttt{jaxoplanet} software.}
\end{figure*}

We randomly select KOI 247.01 as the test planet. We process the \textit{Kepler} light curve using the process described in \citetalias{sagear_orbital_2023}. Then, we fit the processed light curve with \texttt{jaxoplanet} using the procedure and free parameters described in Section \ref{sec:analysis}. We compare the resulting eccentricity posterior to the KOI 247.01 eccentricity posterior from \citetalias{sagear_orbital_2023} in Figure \ref{fig:koi247_exojax}. We demonstrate that we obtain consistent $e$--$\omega$ combined posteriors with each method. Due to the nature of the importance sampling technique (namely, the up-weighting of uniformly spaced test $e$ and $\omega$ arrays resulting in fewer distinct $e$ and $\omega$ posterior values than directly sampling the parameters), the importance-sampled posterior may not appear as ``smooth" as the directly sampled posterior from \citetalias{sagear_orbital_2023}; however, given that we use enough distinct $e$ and $\omega$ test values to populate the ranges where these variables are defined, this has no practical effect on the resulting $e$ and $\omega$ constraints.

\subsection{\textit{TESS} vs. \textit{Kepler} Baseline Observation}

The \textit{TESS} baseline observation time is generally much shorter than the \textit{Kepler} baseline observation time. \textit{TESS} targets are observed for a minimum of 27 days and a maximum of 351 days for targets within the Continuous Viewing Zone \citep{stassun_tess_2018}, while most \textit{Kepler} targets are observed continuously for multiple years \citep{borucki_kepler_2010}. However, many \textit{TESS} light curves are recorded with a two-minute cadence, while the majority of \textit{Kepler} transit light curves consist of long cadence data (30 minutes) with a minority of KOIs having some portion of short cadence (one minute) data. We perform a test to compare the eccentricity resolution for a KOI with \textit{Kepler} long-cadence light curves from \citetalias{sagear_orbital_2023} and a TOI with two-minute cadence data.

\begin{figure*}\label{fig:toi_koi_compare}
  \centering
  \includegraphics[width=\linewidth]{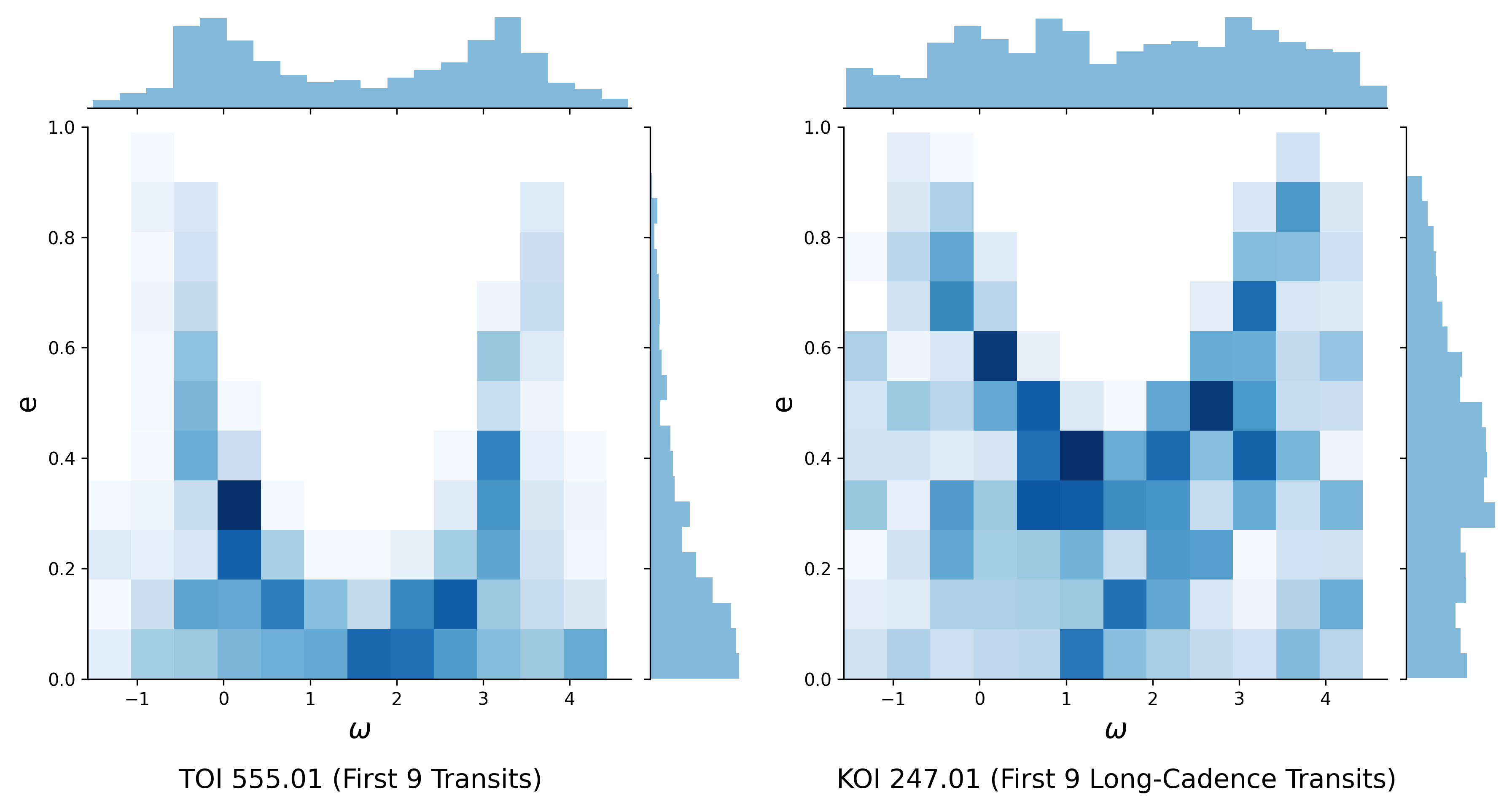}
  \caption{\textit{Left:} Eccentricity posterior generated by fitting the first nine chronological transits of TOI 555.01. \textit{Right:} Eccentricity posterior generated by fitting the first nine chronological long-cadence transits of KOI 247.01.}
\end{figure*}

We begin by selecting a KOI and TOI with similar transit signal-to-noise (SNR). Assuming only the presence of white noise, the SNR of a single transit is defined as

\begin{equation}
    SNR = \frac{\delta}{\sigma_n} \times \sqrt{N}
\end{equation}

where $\delta$ is the transit depth, $\sigma_n$ is the white noise level of the data and $N$ is the number of data points in-transit \citep{pont_effect_2006}. We select KOI 247.01 from the sample of \citetalias{sagear_orbital_2023}, whose \textit{Kepler} long-cadence transit lightcurve has a single-transit SNR of $\approx 8.3$, and TOI 555.01 from the sample of this work, whose \textit{TESS} lightcurve has a single-transit SNR of $\approx 7.4$. For both test planets, we consider only the first 9 transits from their respective lightcurves. We process and fit the shortened light curves according to the process described in Section \ref{sec:analysis}, and use the importance sampling method to generate the $e$--$\omega$ posteriors. We show both $e$--$\omega$ posteriors in Figure \ref{fig:toi_koi_compare}.

Though the baseline observation time is much shorter for \textit{TESS} compared to \textit{Kepler}, the resulting eccentricity constraints are similar. This is likely because the far shorter \textit{TESS} light curve cadence (2 minutes vs. 30 minutes), and though the \textit{Kepler} light curves have longer baseline observation times, the 30-minute cadence does not resolve the transit ingress/egress shape and transit duration nearly as well. Indeed, \textit{TESS} planet parameters are similarly constrained to \textit{Kepler} planet parameters, and we find that the eccentricity constraints are generally comparable.

\subsection{Individual Planet Eccentricity Posteriors}

\begin{figure*}
    \centering
    \includegraphics[width=\linewidth]{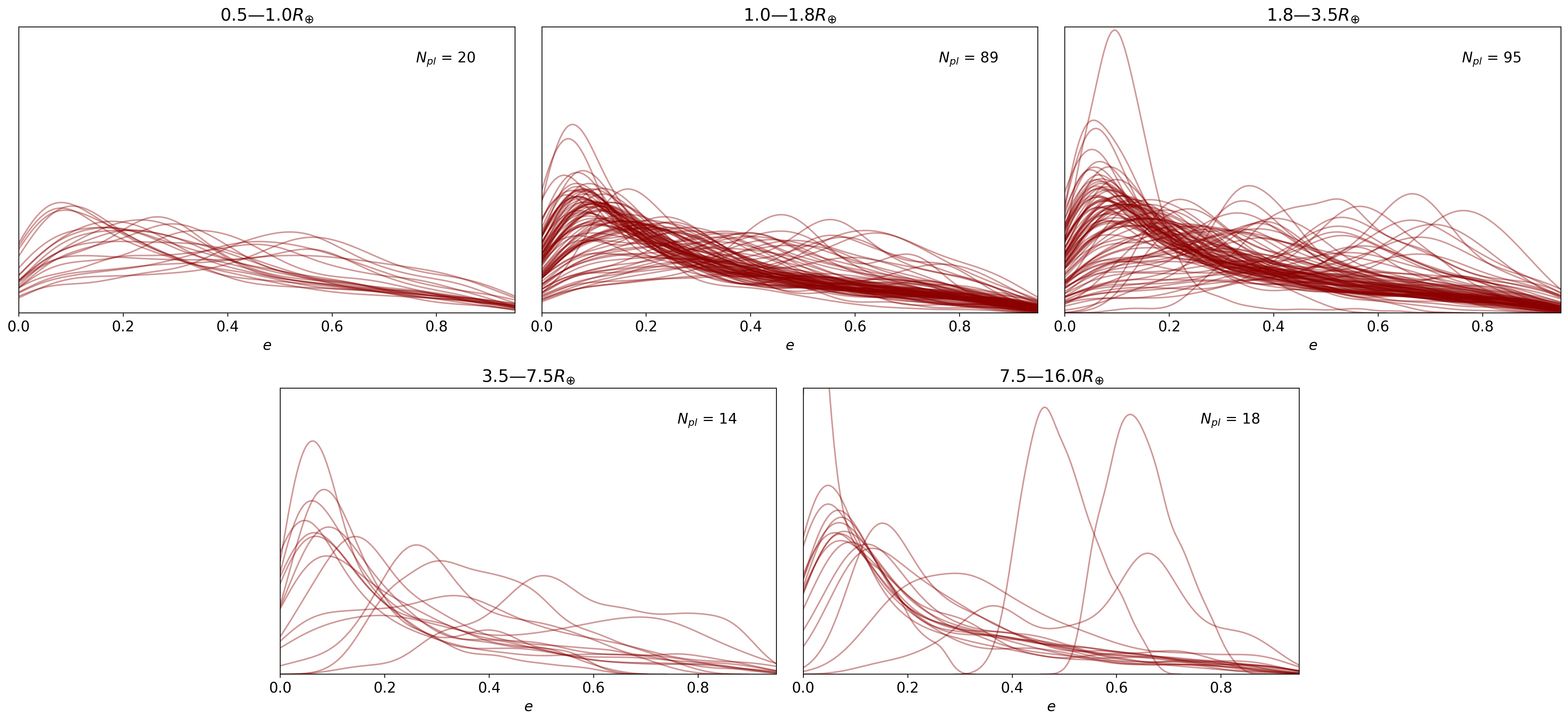}
    \caption{Kernel Density Estimates (KDEs) for the planets in each of five radius bins.}
    \label{fig:KDEs}
\end{figure*}

We present individual planet eccentricity posteriors represented as kernel density estimates (KDEs) in each of the five coarse radius bins in Figure \ref{fig:KDEs}.

\section{Empirical Eccentricity Distribution Model} \label{appendix:empirical}

We detail the process of constructing an empirical eccentricity model by fitting a flexible histogram model to the sample of eccentricity posteriors. We model $p(obs | \theta)$ as a step function (or histogram) with parameters $\theta_m$, a set of $M$ logarithmic bin heights with individual widths $\Delta e_m$:

\begin{equation}
    p{(e|\theta)} = \sum_{m=1}^{M}\exp(\theta_m) f(e; e_{min}+(m-1)\Delta e, e_{min}+m\Delta e)
\end{equation}

where

\begin{equation}
    f(e;a,b) =    
    \begin{cases}
        0, & \text{if } e \le a \text{ or } e \ge b\\
        \frac{1}{b-a}, & a \leq e \leq b
    \end{cases}
\end{equation}

and where $\Delta e = \frac{e_{max}-e_{min}}{M}$. We use $M = 17$ bins, setting individual bin widths such that the $K \times N$ eccentricity samples are evenly divided within $M$ bins, and draw each bin height parameter $\theta_m$ from a normal distribution.

We ensure smoothness between consecutive bin heights by applying a Gaussian Process (GP) prior on the bin heights $\theta$ using a Matérn-3/2 kernel, whose covariance function is

\begin{equation}
    \kappa(e) = s^2 \biggl( 1 + \frac{\sqrt{3}e}{\ell} \biggr) \exp \biggl(- \frac{\sqrt{3}e}{\ell} \biggr)
\end{equation}

where $s$ is the scale and $l$ is the correlation length. As \citetalias{gilbert_planets_2025}, we apply log-normal priors on the GP parameters $\ln(s) \sim \mathcal{N}(3,1)$ and $\ln(\ell) \sim \mathcal{N}(0,1)$.

\begin{figure}
    \centering
    \includegraphics[width=\linewidth]{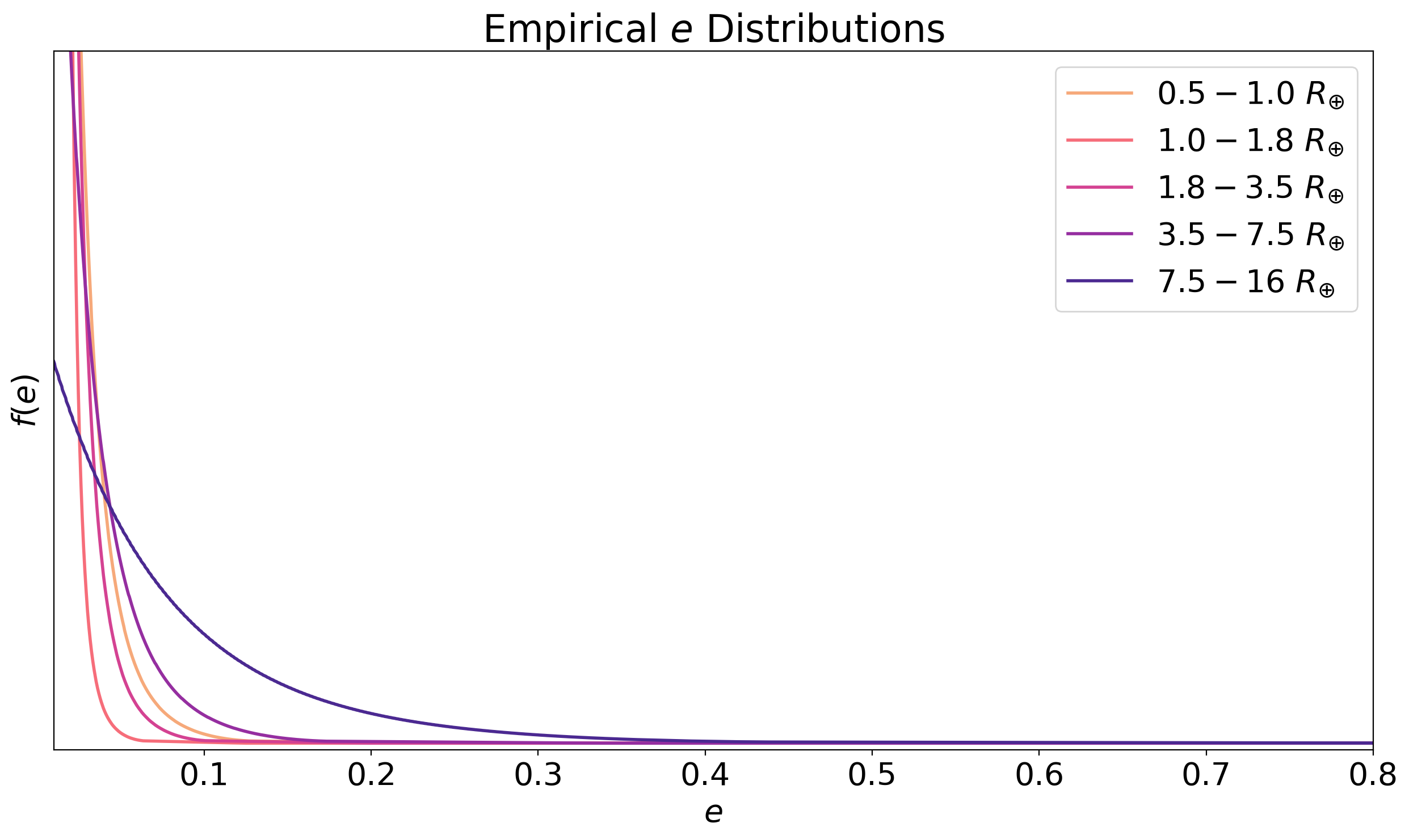}
    \caption{Median empirical eccentricity models for planets in each of the five radius bins in Table \ref{tab:rad_bin_edges}.}
    \label{fig:empirical_distributions}
\end{figure}

\begin{figure}
    \centering
    \includegraphics[width=\linewidth]{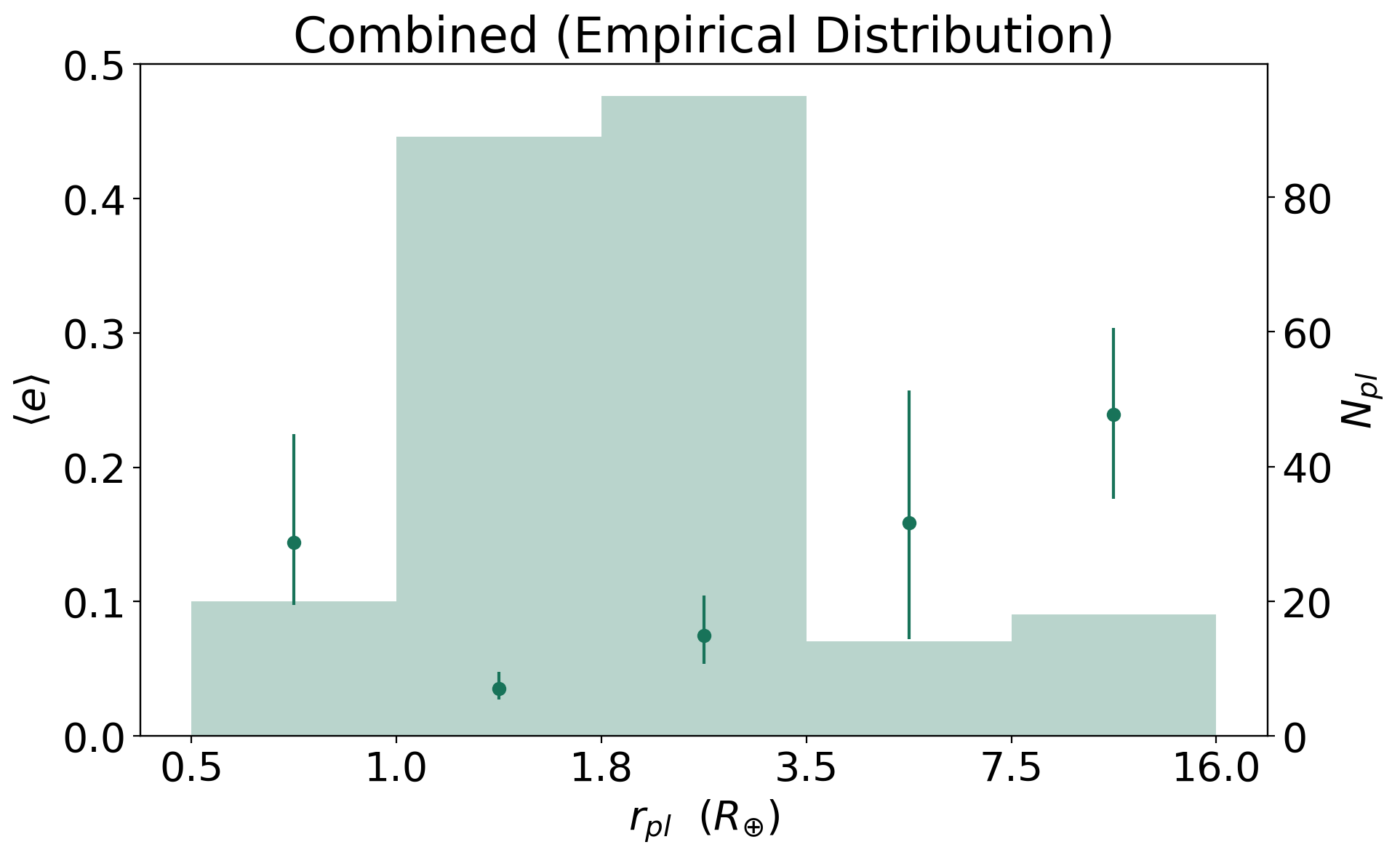}
    \caption{Same as Figure \ref{fig:eccradius_linear}, but where \avge is calculated using the empirical eccentricity model rather than the Beta distribution. The error bars represent the $16^{th}$ and $84^{th}$ percentiles of $\langle e \rangle$. The green histogram represents the number of M dwarf planets in each radius bin (right-hand vertical axis).}
    \label{fig:empirical_bins}
\end{figure}

We implement our regularized histogram model using \texttt{numpyro} \citep{phan_composable_2019} with the GP parameters implemented with \texttt{celerite} \citep{foreman-mackey_fast_2017}. We sample the model parameters (bin heights $\theta$, GP scale $s$ and GP length $\ell$) using the NUTS sampler \citep{homan_no-u-turn_2014} for 2500 tuning steps and 2500 posterior draws. In Figure \ref{fig:empirical_distributions}, we present the median empirical eccentricity distribution for each of the five coarse radius bins (Table \ref{tab:rad_bin_edges}). In Figure \ref{fig:empirical_bins}, we present an analog of Figure \ref{fig:eccradius_linear} using our empirical model instead of the Beta distribution.

\bibliography{main}{}
\bibliographystyle{aasjournalv7}

\end{document}